\documentclass[fleqn,usenatbib]{mnras}

\usepackage{newtxtext,newtxmath}

\usepackage[T1]{fontenc}
\usepackage{ae,aecompl}

\usepackage{graphicx}	
\usepackage{natbib}
\usepackage{soul}
\bibpunct{(}{)}{;}{a}{}{,}
\usepackage{graphicx}
\usepackage{multirow}

\usepackage{graphicx}
\usepackage{txfonts}
\usepackage{natbib}
\usepackage{xcolor}
\usepackage{xspace}
\usepackage{dcolumn}
\usepackage{longtable}
\usepackage{lscape}
\usepackage{soul}        

\usepackage{rotating}
\usepackage{afterpage}

\definecolor{amethyst}{rgb}{0.6, 0.4, 0.8}

\newcommand{\MJup}{M$_{\mathrm{Jup}}$\xspace}

\newcommand{\mic}{$\mu$m\xspace}

\title[Stellar multiplicity in Lupus.]{The effect of stellar multiplicity on protoplanetary discs: \\ A NIR survey of the Lupus star forming region. }

\author[A. Zurlo et al.]{Alice Zurlo$^{1,2}$\thanks{E-mail: alice.zurlo@mail.udp.cl}, Lucas A. Cieza$^{1}$, Megan Ansdell$^{3}$, Valentin Christiaens$^{4}$, \newauthor Sebasti\'an P\'erez$^{5}$, Josh Lovell$^{6}$, Dino Mesa$^{7}$, Jonathan P. Williams$^{8}$, \newauthor Camilo Gonzalez-Ruilova$^{1}$, Rosamaria Carraro$^{9}$, Dary Ru\'iz-Rodr\'iguez$^{10}$, Mark Wyatt$^{5}$ \\ \\
$^{1}$N\'ucleo de Astronom\'ia, Facultad de Ingenier\'ia, Universidad Diego Portales, Av. Ejercito 441, Santiago, Chile\\
  $^{2}$Escuela de Ingenier\'ia Industrial, Facultad de Ingenier\'ia y Ciencias, Universidad Diego Portales, Av. Ejercito 441, Santiago, Chile\\
$^{3}$ National Aeronautics and Space Administration Headquarters, 300 E Street SW, Washington, DC 20546, USA\\
$^{4}$School of Physics and Astronomy, Monash University, Clayton VIC 3800, Australia\\
$^{5}$Departamento de F\'isica, Universidad de Santiago de Chile, Av. Ecuador 3493, Estaci\'on Central, Santiago, Chile \\
$^{6}$Institute of Astronomy, University of Cambridge, Madingley Road, Cambridge CB3 0HA, UK\\
$^{7}$INAF - Osservatorio Astronomico di Padova, Vicolo dell'Osservatorio 5, 35122, Padova, Italy \\
$^{8}$Institute for Astronomy, University of Hawaii at Manoa, Honolulu, HI, 96822, USA\\
$^{9}$Instituto de F\'isica y Astronom\'ia, Universidad de Valpara\'iso, Gran Breta\~na 1111, Playa Ancha, Valpara\'iso, Chile \\
$^{10}$Chester F. Carlson Center for Imaging Science, School of Physics \& Astronomy, and Laboratory for Multiwavelength Astrophysics,\\
  Rochester Institute of Technology, 54 Lomb Memorial Drive, Rochester NY 14623 USA
}
\date{ }

\pubyear{2020}

\begin{document}
\label{firstpage}
\pagerange{\pageref{firstpage}--\pageref{lastpage}}
\maketitle

\begin{abstract}
{We present results from a near-infrared (NIR) adaptive optics (AO) survey of pre-main-sequence stars in the Lupus Molecular Cloud with VLT/NACO to identify (sub)stellar companions down to { $\sim$20} au separation and investigate the effects of multiplicity on circumstellar disc properties. We observe for the first time in the NIR with AO a total of 47 targets and complement our observations with archival data for another 58 objects previously observed with the same instrument. All 105 targets have millimetre ALMA data available, which provide constraints on disc masses and sizes. We identify a total of { 13} multiple systems, including { 11} doubles and { 2} triples. In agreement with previous studies, we find that the most massive (M$_{\rm dust}$ $>$ 50 M$_{\oplus}$) and largest ($R_{\rm dust}>$ 70 au) discs are only seen around stars lacking visual companions (with separations of { 20--4800} au) and that primaries tend to host more massive discs than secondaries. However, as recently shown in a very similar study of $>$200 PMS stars in the Ophiuchus Molecular Cloud, the distribution of disc masses and sizes are similar for single and multiple systems for M$_{\rm dust} < 50$ M$_{\oplus}$ and radii $R_{\rm dust}<$ 70 au. Such discs correspond to $\sim $80--90\% of the sample. This result can be seen in the combined sample of Lupus and Ophiuchus objects, which now includes more than 300 targets with ALMA imaging and NIR AO data, and implies that stellar companions with separations $>${ 20} au mostly affect discs in the upper 10$\%$ of the disc mass and size distributions.}
\end{abstract}




\section{Introduction}

Recent Atacama Large Millimetre/sub-millimetre Array (ALMA) surveys in nearby ($d\!<\!300$~pc) molecular clouds such as Lupus and Ophiuchus \citep{Ansdell2016, Cieza2019} have shown that protoplanetary discs exhibit a wide range of masses ($<$\,1--200 M$_{\oplus}$) and sizes (r $<$ 15--300 au). Disc properties are a function of different variables, including infrared spectral energy distribution (SED) class \citep{Williams2019}, stellar mass \citep{Andrews2013, Pascucci2016, Barenfeld2016}, age \citep{Ansdell2018, RuizRodriguez2018}, and multiplicity \citep{Harris2012, Cox2017, Zurlo2020}. Disentangling the effects of each variable is difficult and requires large samples and multi-wavelength observations. 

In general, studies have found that discs in binary systems with medium separations ($\sim$10--200~au) tend to be smaller and less massive than those around single stars. Very tight systems (with separations $<$1~au) seem to host massive circumbinary discs, while wide companions (with separations $>$200~au) have little impact on disc properties \citep{Harris2012, Cox2017}. { These effects that stellar companions have on disc properties can be understood in terms of tidal truncation. In the case of a circumbinary disc, the inner edge location varies from 1.8--2.6 times the separation, depending on the eccentricity of the orbit \citep[e.g.][]{1994ApJ...421..651A}. For a system with a stellar companion external to the circumstellar disc, truncation occurs at 0.3--0.5 times the physical separation of the binary companion \citep[see, for example,][]{Papaloizou1977, 1994ApJ...421..651A}.}\footnote{These truncation radii rely on co-planarity of the discs with the binary orbit and could change if more complex scenarios are considered.} A smaller disc also implies a shorter viscous dissipation timescale and previous infrared studies have shown that discs around medium-separation binaries do in fact have shorter lifetimes than those around single stars { \citep{Cieza2009, Kuruwita2018}}. In addition, \citet{2018MNRAS.473.5630R} found that secondary components of close binaries clear first due to the shorter viscous time scale associated with the smaller size of the disc, while for wide binaries the difference in photoevaporation rate permits a longer lifetime than tight systems. 

Nevertheless, the extent to which stellar multiplicity affects disc properties is not yet fully established. For instance, \citet{Cox2017} found significant differences in the disc sizes and masses seen around binary systems with respect to those of single stars. However, as noted by \citet{Zurlo2020}, their sample was selected based on the presence of 70\,$\mu$m \emph{Spitzer} excesses, which introduces strong biases against small and faint discs. Since the vast majority of discs in Ophiuchus are very small \citep[r\,$<$\,15~au,][]{Cieza2019} and low-mass \citep[M$_{\rm dust}$\,$\lesssim$\,1~M$_\oplus$,][]{Williams2019}, the effects that visual binaries (with separations { $>$\,20~au}) have on the sizes and masses of the general disc population are weak \citep{Zurlo2020}.  

In order to investigate disc properties as a function of multiplicity, we present here a near-infrared (NIR) adaptive optics (AO) survey with VLT/NACO for a sample of 125 young stellar objects (YSOs) in the Lupus star-forming region, which have previously been observed by ALMA at $\sim$0\farcs2 resolution for the Class II objects \citep{Ansdell2016, Ansdell2018} and at $\sim$0\farcs7 resolution for the Class III objects \citep{2020MNRAS.tmp.3151L}. The sample selection of this NIR survey is presented in Section~\ref{s:sample}. The observations and data reduction of the NIR and millimetre data are described in Section~\ref{s:obs}. Our main results combining the NACO and ALMA data are presented in Section~\ref{s:res}.  In Section~\ref{s:discussion} we discuss our results, comparing and combining them with those from other regions. Finally, we present a summary of our conclusions in Section~\ref{s:con}.

\section{The Lupus Sample and literature data}
\label{s:sample}

The aim of this NIR survey is to study the multiplicity of YSOs in the Lupus star-forming region and the effects on circumstellar disc properties. The NIR-Lupus sample presented in this paper consists of the YSOs more massive than brown dwarfs in the Lupus I-IV clouds that have been identified as either Class~II or Class~III sources. The complete sample includes 125 objects: 95 Class II sources that were observed with ALMA and presented in \citet{Ansdell2016, Ansdell2018} and 30 Class III objects that were observed with ALMA and { presented in \citet{2020MNRAS.tmp.3151L}}. 

Among the 125 objects in our NIR-Lupus sample, we found NACO observations for 58 of them in the European Southern Observatory (ESO) archive. A total of 47 objects have never been observed in the NIR with AO and are bright enough to be observed with NACO, with a $K$ magnitude of 12.6 for the faintest target. These 47 objects were included in our NACO target list, along with the stars in our sample with poor-quality archival data and all the known binaries that we were allowed to re-observe with NACO. There are 20 objects in our sample that are too faint for NACO observations; these objects are listed in Table~\ref{t:non_o} and are excluded from our statistical analysis. 

The literature covering multiplicity in Lupus is currently limited. Some visual binaries were presented in \citet{Reipurth1993}: Sz~65 (the companion to Sz~66), Sz~68, Sz~81, Sz~88, and Sz~108.  Sz~65 was also identified as a binary by \citet{Merin2008}, while Sz~69, presented in the same paper as a binary, appears as a single star in our survey. The discs in the system of Sz 68 (also known as HT Lup) have been resolved by ALMA \citep[as part of DSHARP,][]{2018ApJ...869L..41A,2018ApJ...869L..44K} and in polarized scattered light with SPHERE \citep[as part of DARTTS-S,][]{2020A&A...633A..82G}. Sz~74 was also identified as a binary system by \citet{Woitas2001} and Sz~123 is composed of two close M-type star binaries as reported by \citet{alcala2014}.

\section{Observations and data reduction}
\label{s:obs}

\subsection{VLT/NACO}
\label{s:obsnaco}

To study the multiplicity in Lupus, we observed the YSOs in our sample with VLT/NACO, using methods similar to the already completed NIR-ODISEA survey of the Ophiuchus molecular cloud \citep[][]{Zurlo2020}. We refer the reader to \citet{Zurlo2020} for details on the observing strategy. In short, we used very quick (0.16--0.17~s) $L^{\prime}$ band filter exposures with the ``star-hopping'' technique to minimize overheads and maximize the number of observed objects. We took 2 exposures for each target with the jittering technique to record a quasi-simultaneous sky background. As for NIR-ODISEA, we chose the AO configuration N90C10 (where 90\% of the star light goes to the AO system NAOS and the remaining 10\% goes to the detector). The observations were carried out in visitor mode during the nights of 6, 20, 22, and 23 of August 2019 (for a total of 1.2 observing nights) under ESO program 0103.C-0466 (PI: Zurlo). The observing conditions were generally favourable, with seeing from 0\farcs4 to 1\farcs2 (averaging 0\farcs6), low wind, and stable observing conditions. { This permitted us to swiftly observe all of the 47 new targets, and to re-observe some of the known multiple systems.}

The data reduction follows the procedures described in \citet{Zurlo2020}: the two exposures per star are background subtracted (one from the other), flat-fielded, and then re-centred to have the target in the center of the detector. Each final image is the mean of the two re-centred frames. For the astrometric calibration, we adopt the ALMA coordinates for the center of the star, the true North direction as given by the position angle of the exposure on sky, and the pixel scale of 0.02719 arcsec/pixel. { The field of view is about 32 arcseconds wide, which is $\sim$4800 au at 150 pc distance.} 

{ The companion detection procedure starts with the automatic identification of  $>$5$\sigma$ signals using the \emph{Daophot} algorithm via the {\tt photutils} package in the {\tt astropy} library \citep{astropy:2018}. All of these automatic detections are inspected visually for confirmation in the individual images (in very few cases some artifacts were selected by the algorithm). The minimum separation probed by our observations is related to the AO diffraction limit of the instrument in $L^{\prime}$ band. This is 0\farcs1, i.e. $\sim$18 au at 150 pc.  Binaries with separations closer than 20 au would not be resolved in this study.  A conservative error of half a pixel is given on the measured separation of the companions, which corresponds approximately to the width of the point spread function (PSF) divided by the minimum signal-to-noise ratio (SNR) of our observations.} Note that an offset of up to 2 degrees on the North orientation may be present, as reported in the NACO manual, version P103. 

The detected multiple systems are shown in Figures~\ref{f:mos} and \ref{f:mos2}. The individual values of all the contrast curves are shown in Figure~\ref{f:contr}. { Contrast limits are calculated {from the} azimuthal median of the rms noise, following \cite{2014A&A...572A..85Z}.}  { Two triple systems and 11 binaries are detected in the final NACO images}. One of the two triple systems is Sz~123, reported as close binary by \citet{alcala2014}, which shows a faint tertiary in the NW. { This detection has the lowest local SNR of $\sim$6.} The properties of the companion sources are listed in Table~\ref{t:bin}.

\subsection{Archival data}

\begin{figure*}

\begin{center}
  \includegraphics[height=0.3\textwidth]{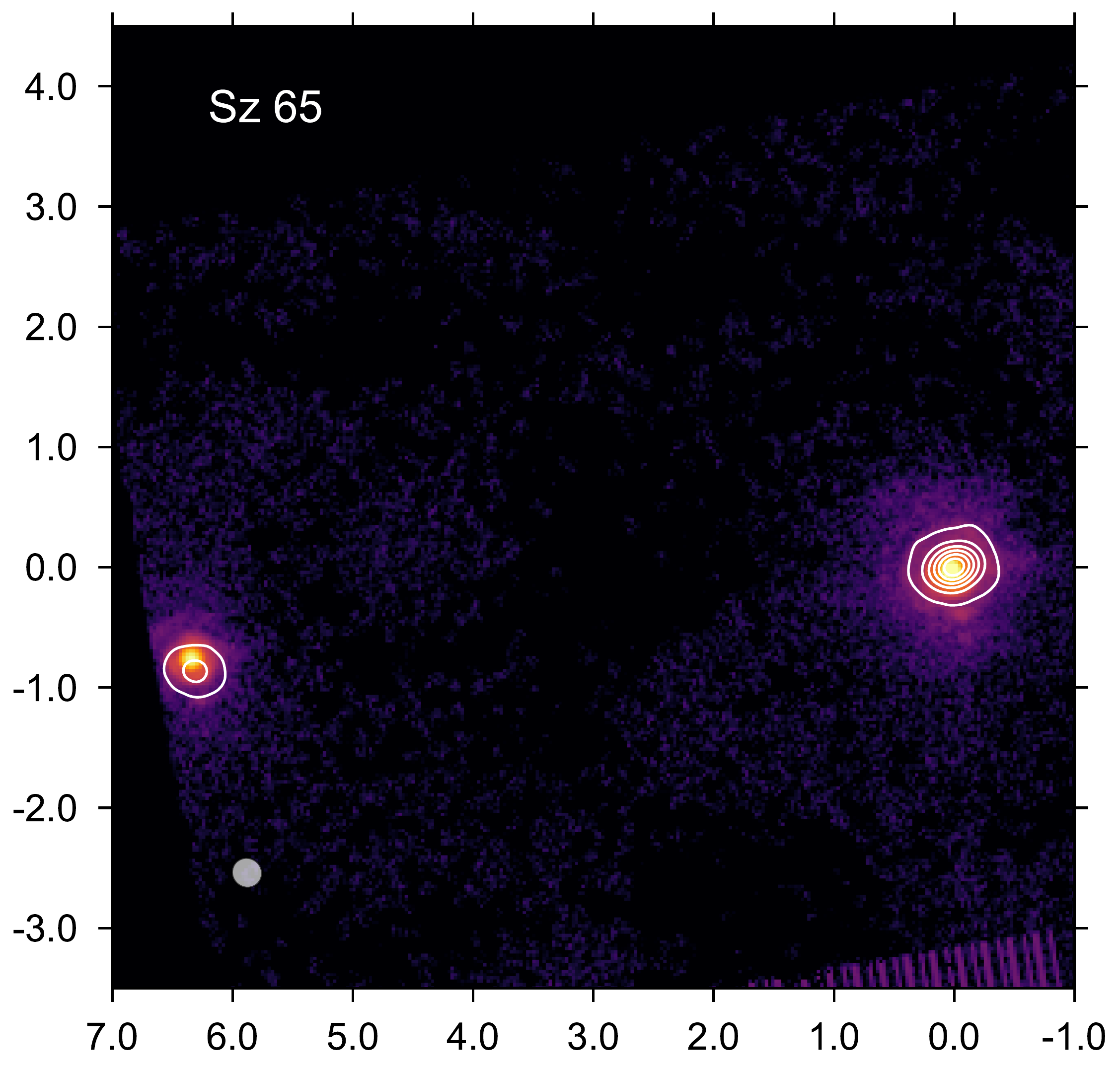}  \hfill
  \includegraphics[height=0.3\textwidth]{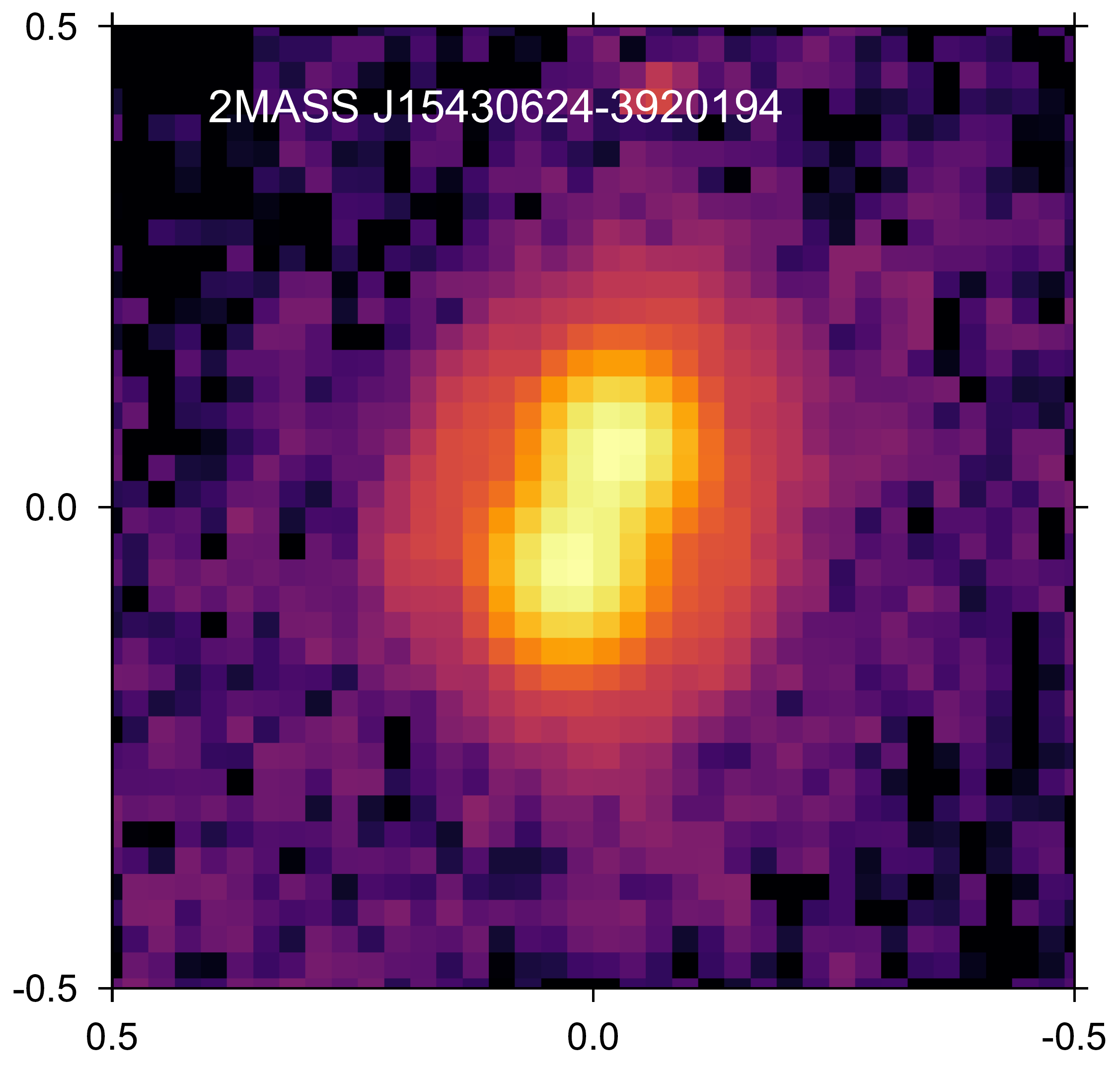}  \hfill
  \includegraphics[height=0.3\textwidth]{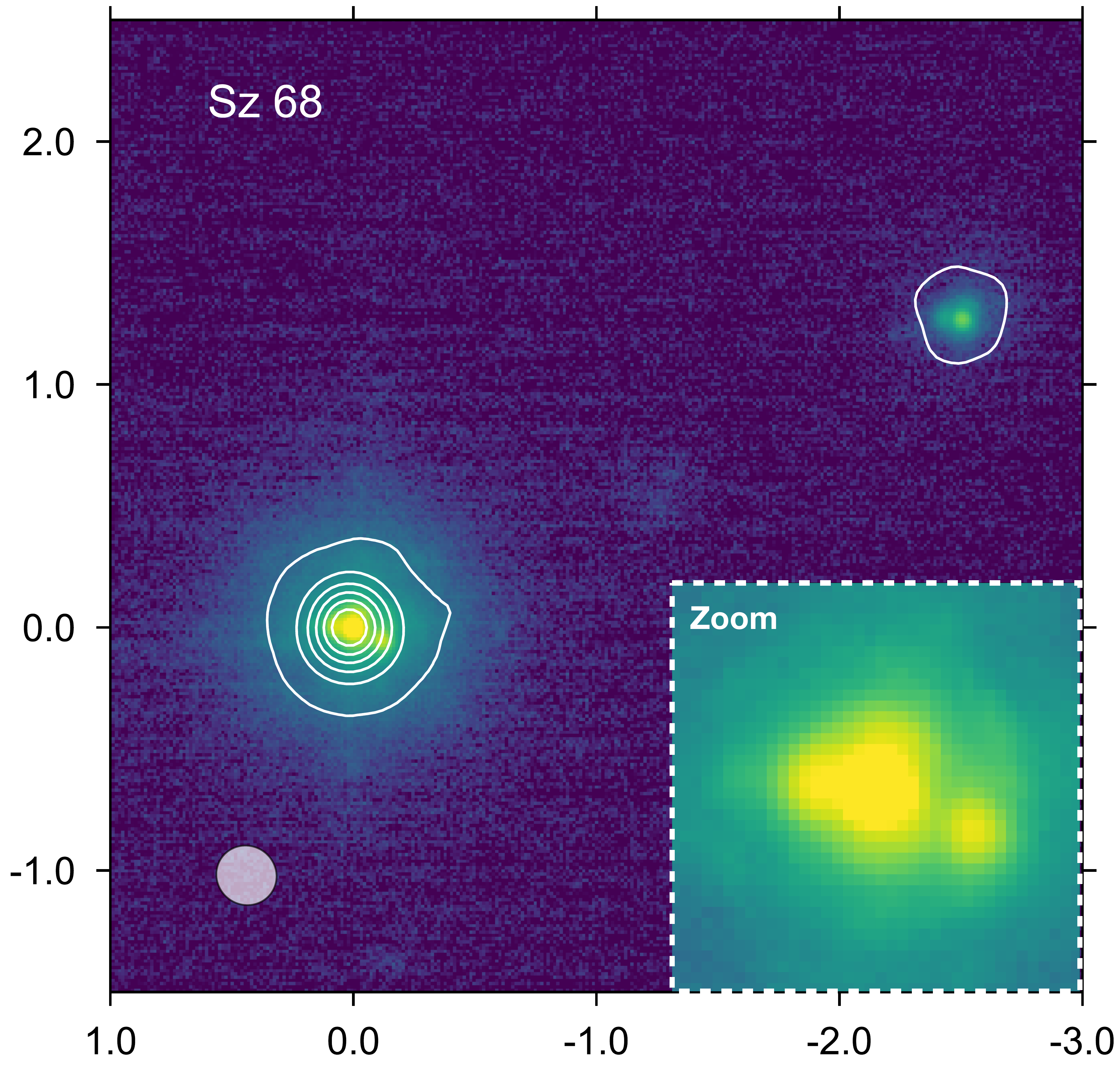}  \\
  \includegraphics[height=0.3\textwidth]{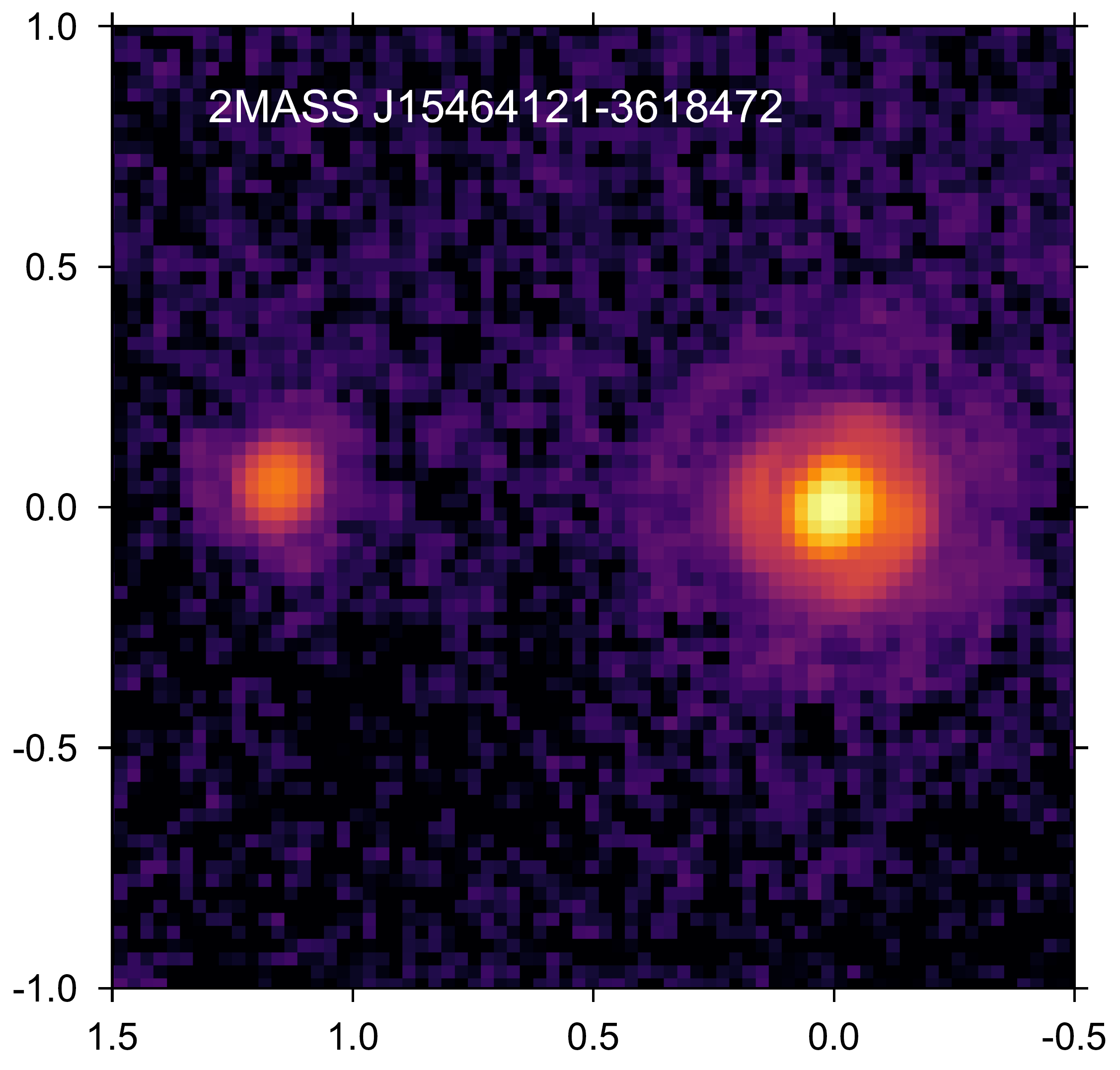} \hfill 
  \includegraphics[height=0.3\textwidth]{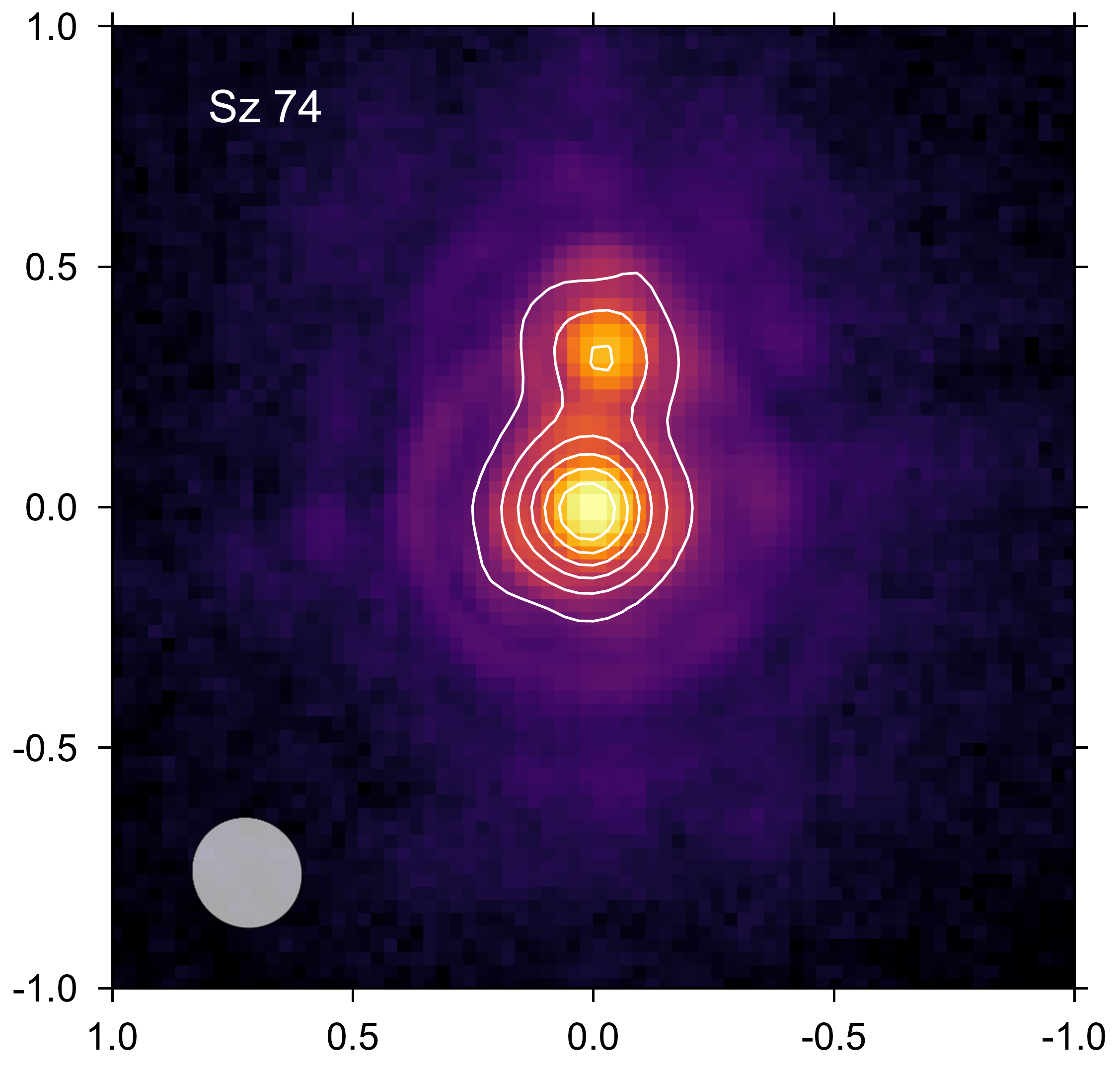}  \hfill
  \includegraphics[height=0.3\textwidth]{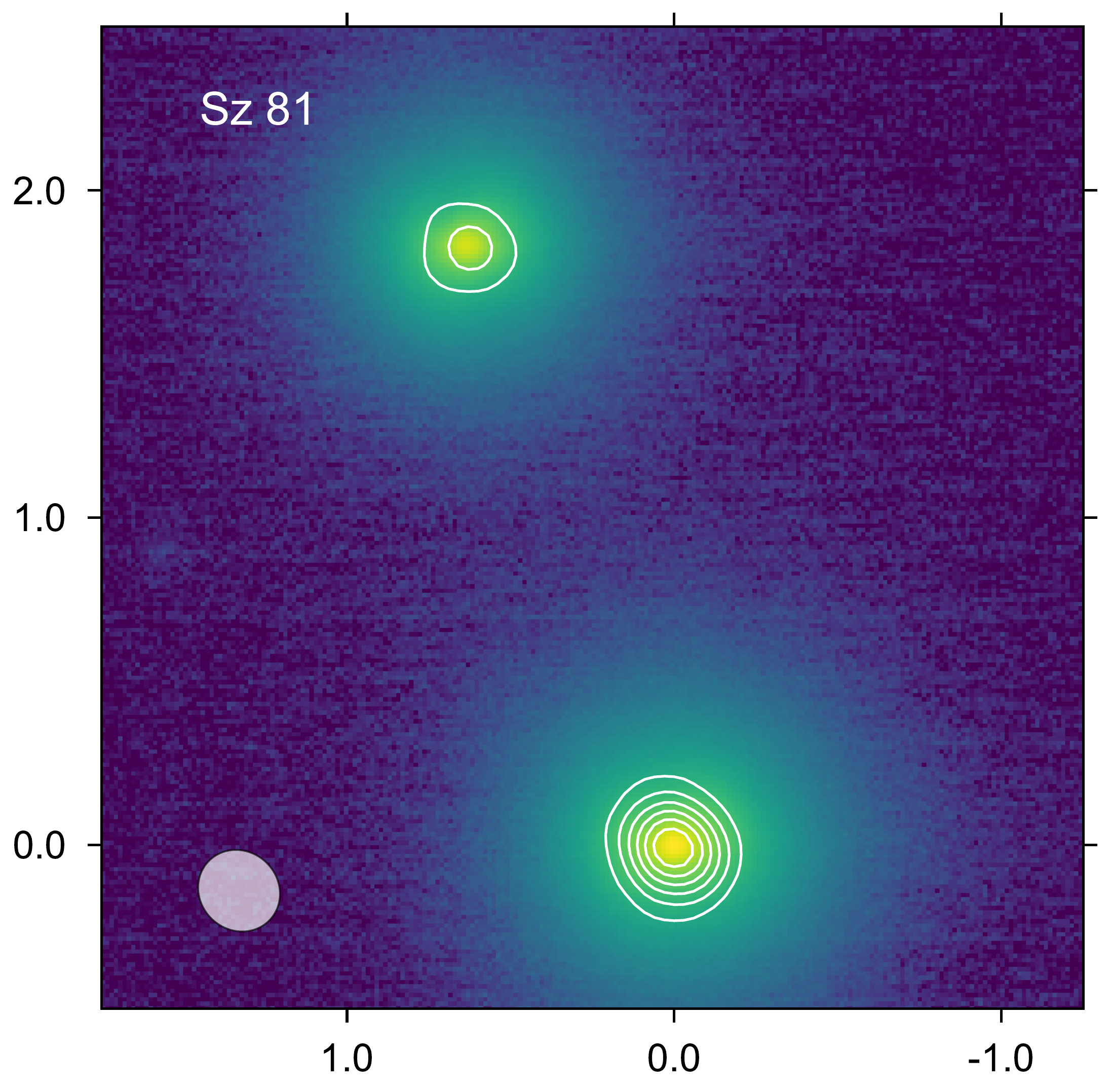}  \\
  \includegraphics[height=0.3\textwidth]{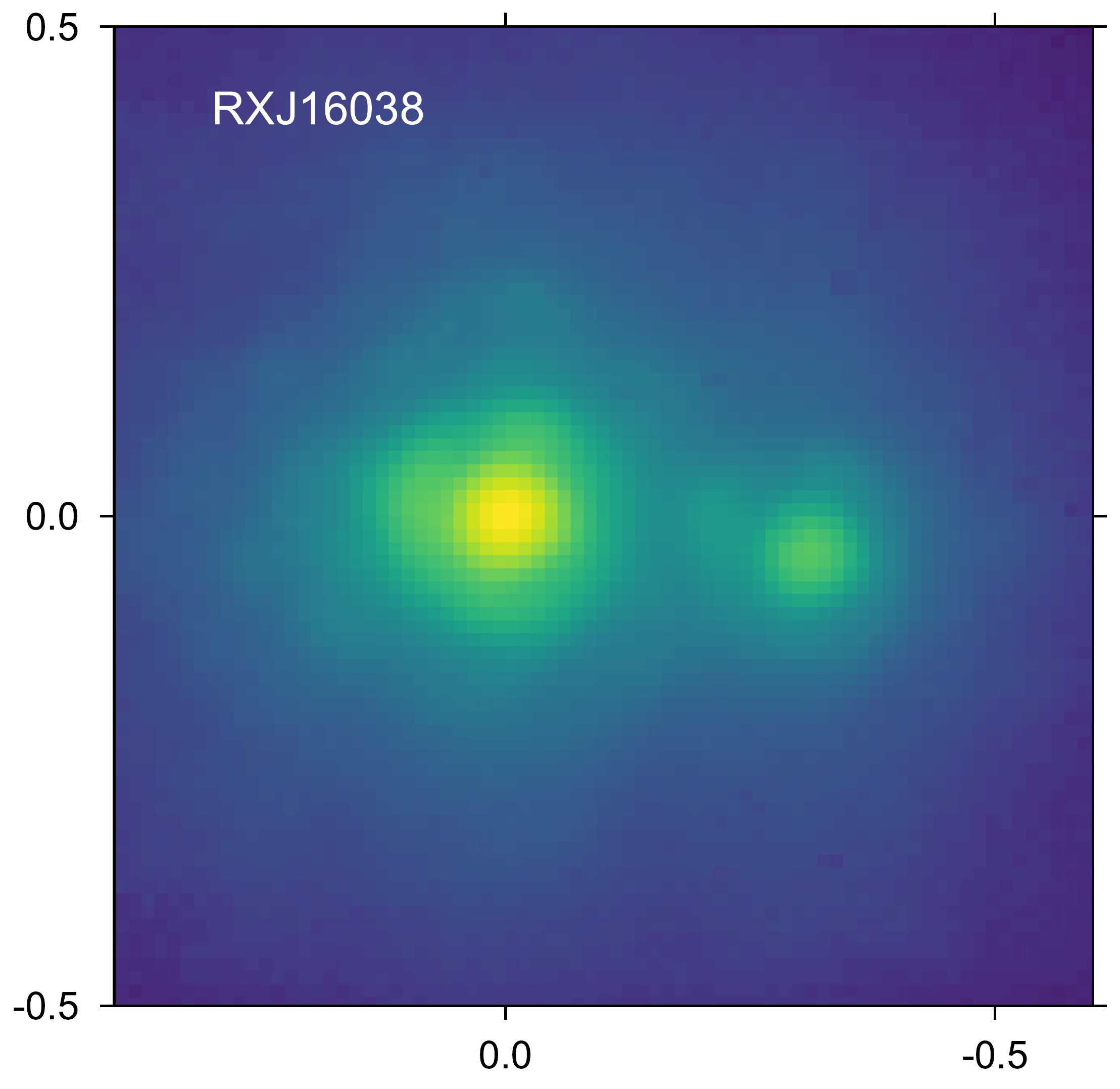}  \hfill
  \includegraphics[height=0.3\textwidth]{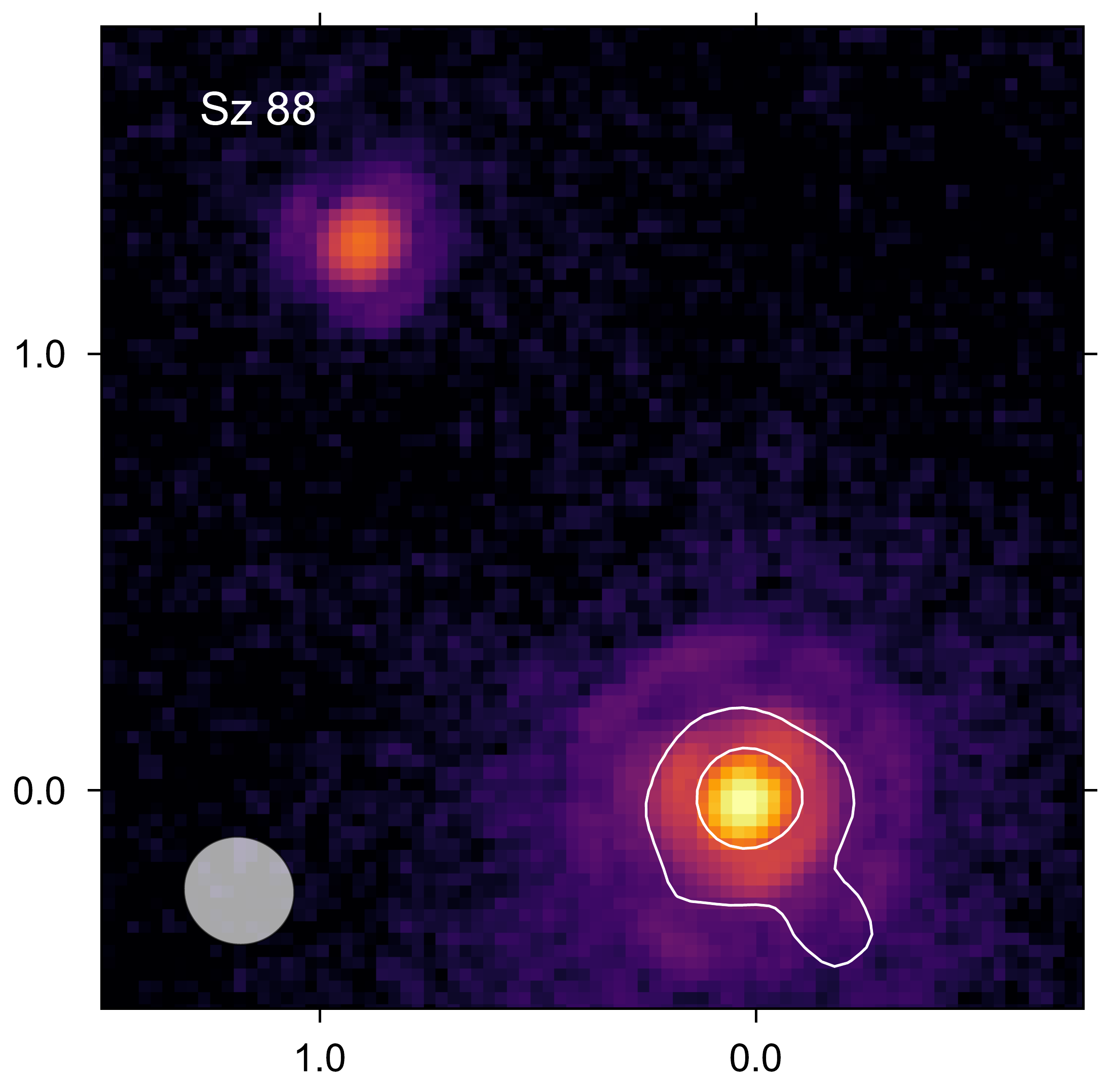}   \hfill
  \includegraphics[height=0.3\textwidth]{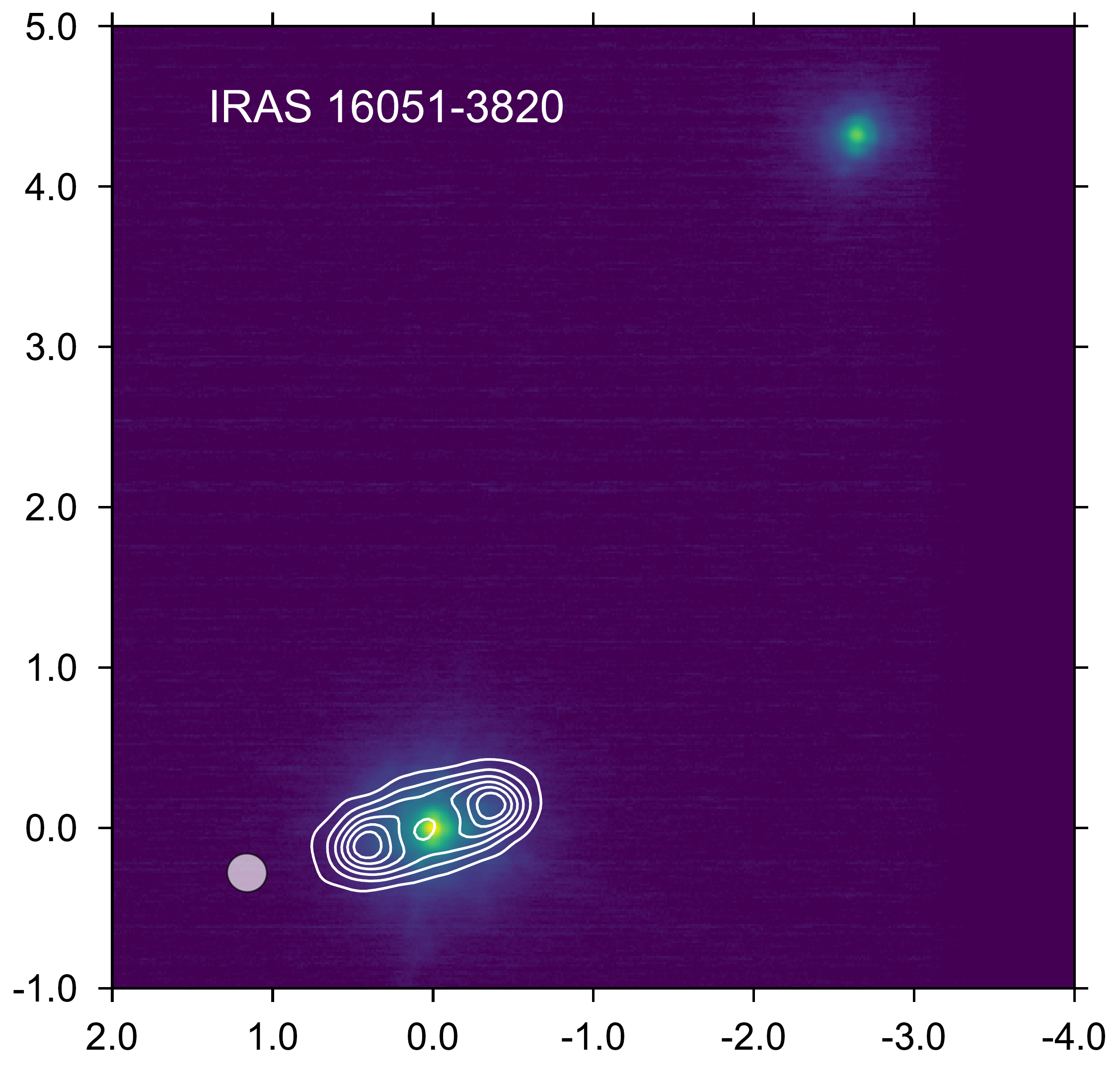}    \\
  \includegraphics[height=0.3\textwidth]{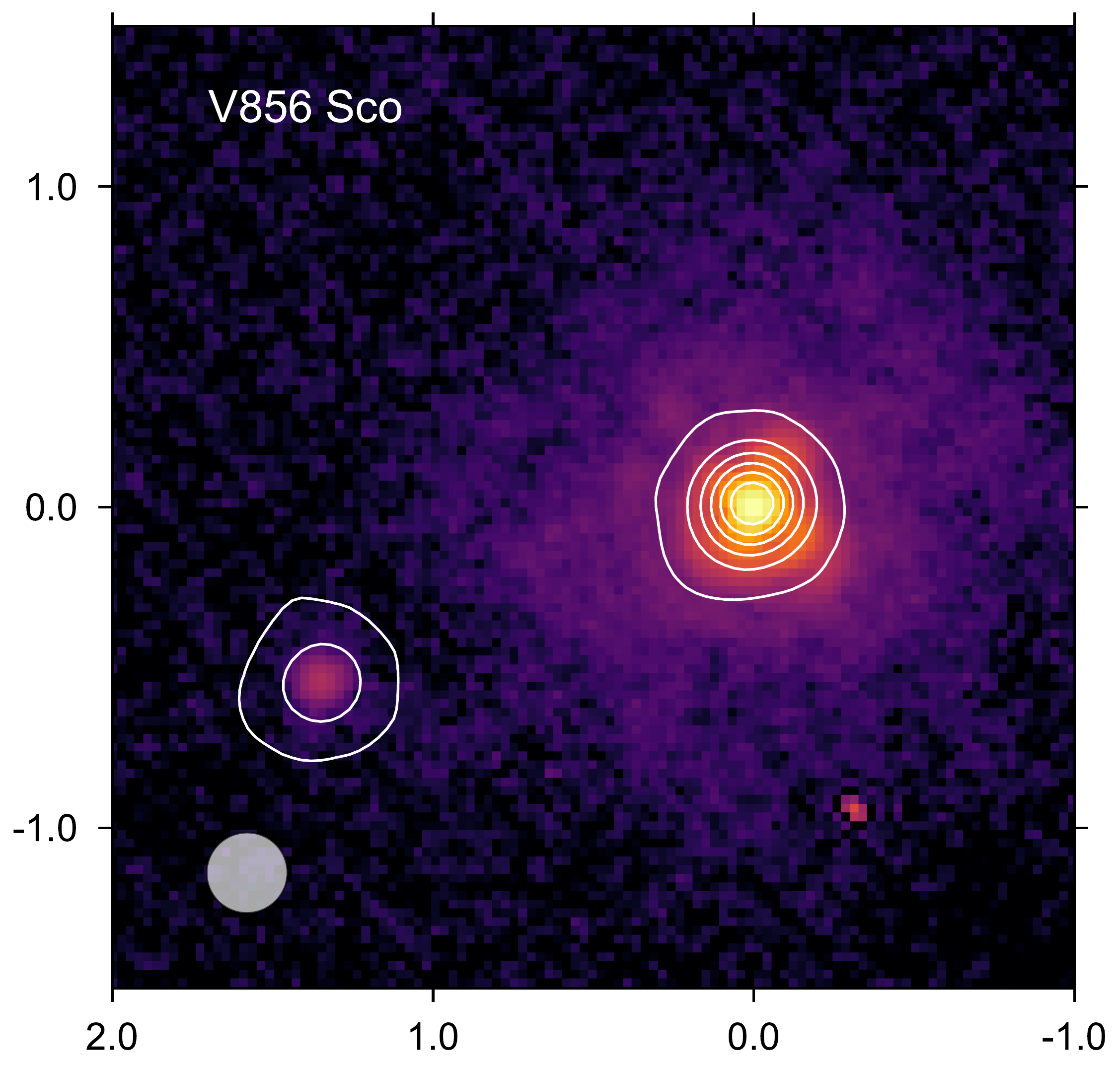}  \hfill
  \includegraphics[height=0.3\textwidth]{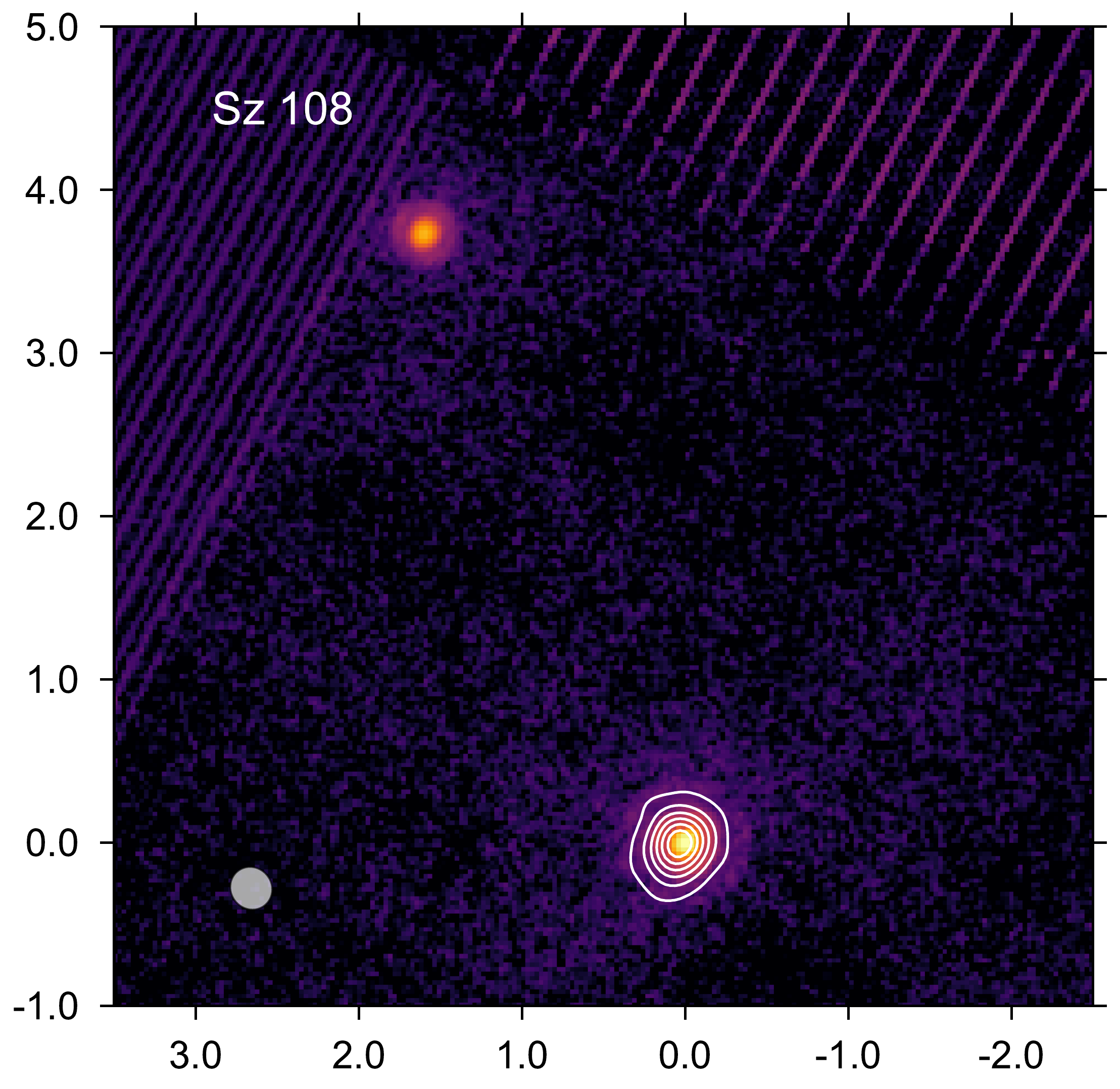}  \hfill
  \includegraphics[height=0.3\textwidth]{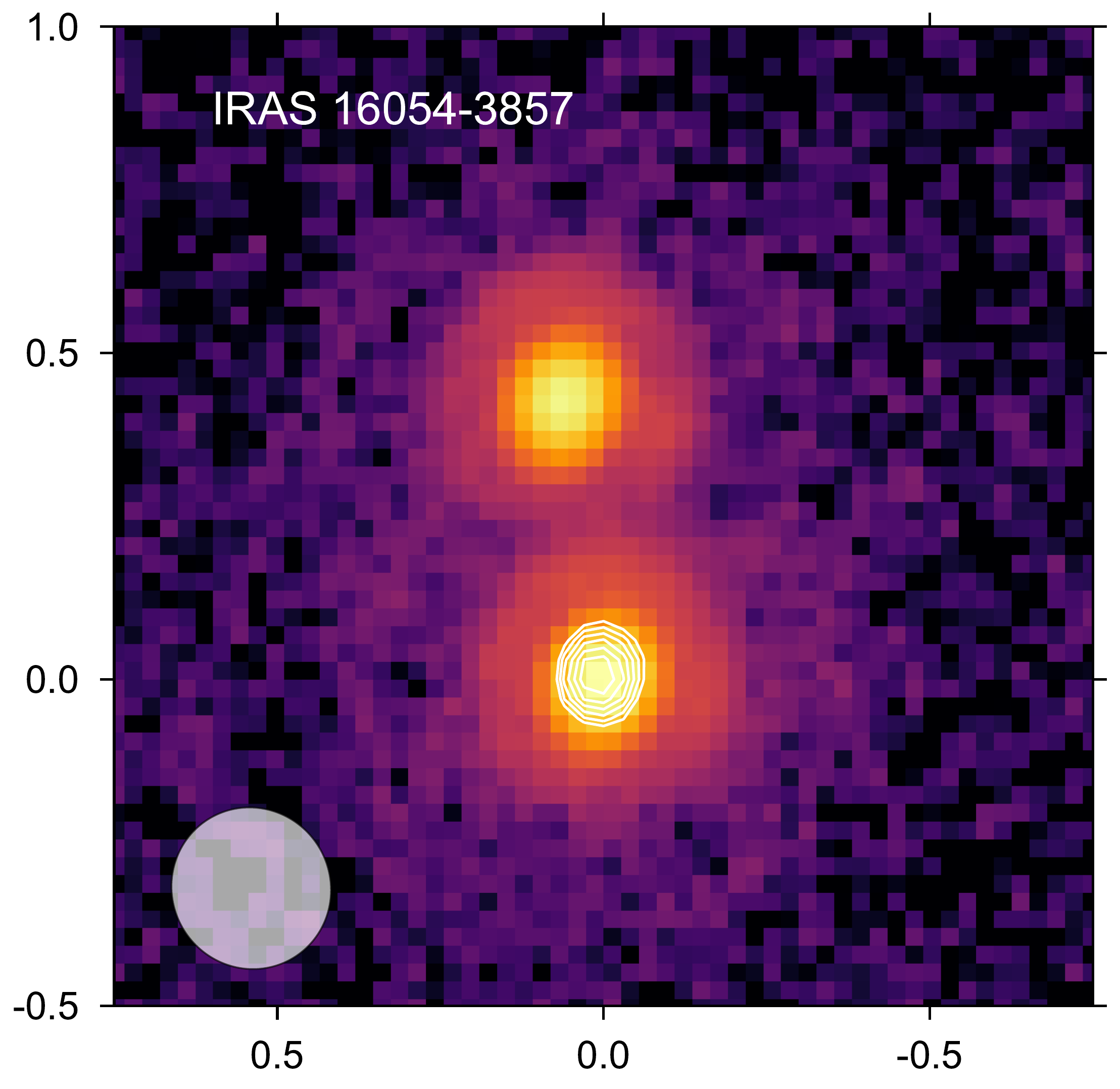}\\

  \caption{A gallery including all the {detected multiple systems} of the Lupus sample. For each object, the NACO image is shown with a logarithmic colour stretch. Images shown with a distinct green colour map correspond to archival data. When detected, the millimetre ALMA counterpart 1.3 mm emission is shown in white contour levels, ranging from 5 times the RMS noise (normally the RMS noise is 0.15-0.2 mJy in each map) to 90\% the peak emission. None of the Class III objects are detected in the mm. The ALMA synthesized beam is shown in the bottom left corner. North is up, East is left, scale in arcsec.}


\label{f:mos}
\end{center}
\end{figure*}

We found 58 objects in our Lupus sample in the ESO archive, all observed with NACO but through different observing programs using different techniques and filters. We reduced all the raw archival images with the same technique described in Section~\ref{s:obsnaco} and included any objects with low-quality or inconclusive data in our NACO target list for an additional epoch of observation. Four systems are multiple: Sz 68 (triple system, observed under program 073.C-0379), Sz 81 (095.C-0610), RXJ16038 (085.C-0012), and IRAS 16051-3820 (097.C-0572). All of them but Sz 68 were observed in $K_{\rm s}$ band with the jittering technique; Sz 68 was observed with the NB\_1.64 filter. The final images of the archival data are shown in Figure~\ref{f:mos} with a different color map to distinguish them from our new NACO observations. { Contrast curves are shown in Figure~\ref{f:contr}}. Two new binaries are found, while Sz~68 is discovered to be a triple system, not a binary as presented in \citet{Reipurth1993}. Note that all the multiple systems reported in  the literature were re-observed in this survey, or reduced from archival data. The properties of the systems are listed in Table~\ref{t:bin}.

\subsection{ALMA data}

The ALMA data for the Class II sources in our Lupus sample were taken during Cycle 3 {in Band~6} under program 2015.1.00222.S (PI: Williams). The complete ALMA dataset {of all 95 Class II sources in Lupus} was presented in \cite{Ansdell2018}. The median {1.3~mm} continuum rms was 0.10 mJy beam${^{-1}}$ and the typical beam size was { $0.25^{\prime\prime}$ ($\sim$40~au at 150~pc)}. {Only 24 of the Class II sources were undetected by ALMA, and these typically had spectral types of M3 or later.} The {candidate secondary sources reported in \cite{Ansdell2018}} but detected only in the ALMA data are not considered as multiple systems in this analysis, as they do not have NIR counterparts. These systems are: Sz~88, which presents a marginally detected elongated PSF in ALMA, J16070384-3911113, which is too faint to be observed with NACO, and J16073773-3921388, which appears as a single star in the NACO image. The properties of these systems as reported by \cite{Ansdell2018} are listed in Table~\ref{t:bin_alma}.

The ALMA data for the Class III objects in our Lupus sample were taken during Cycle~8 in Band~7 under program 2018.1.00437.S (PI: Wyatt) and are presented in \citet{2020MNRAS.tmp.3151L}. This Class III ALMA survey had a moderate resolution of $\sim$ 0\farcs7 and a high continuum sensitivity of $\sim$ 37 $\mu$Jy at 860 \mic, but detected only 4 systems. None of the NIR binaries reported in this work are detected in the ALMA data.


\begin{figure*}
\centering
\includegraphics[height=0.3\textwidth]{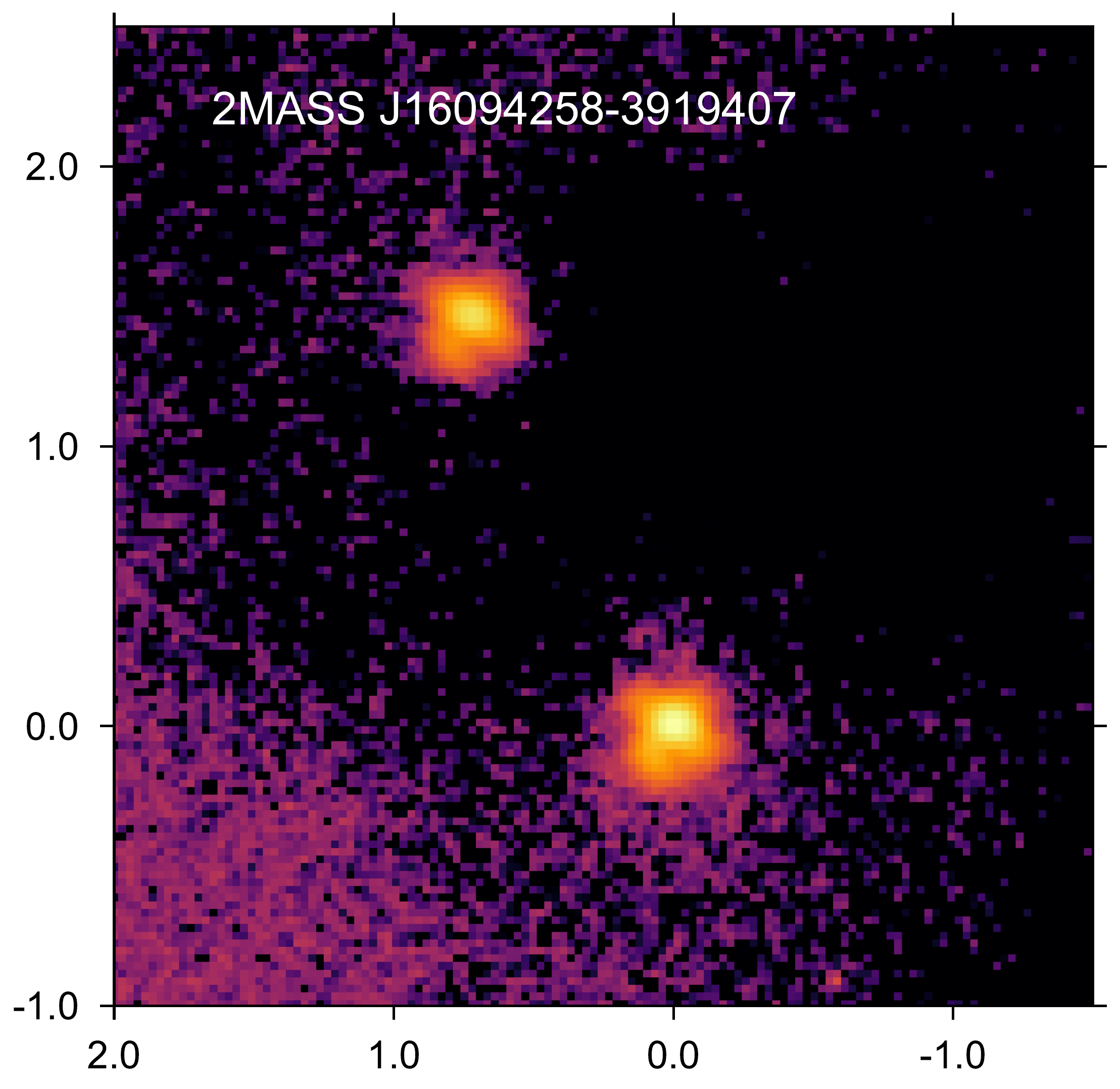} \hfill
\includegraphics[height=0.3\textwidth]{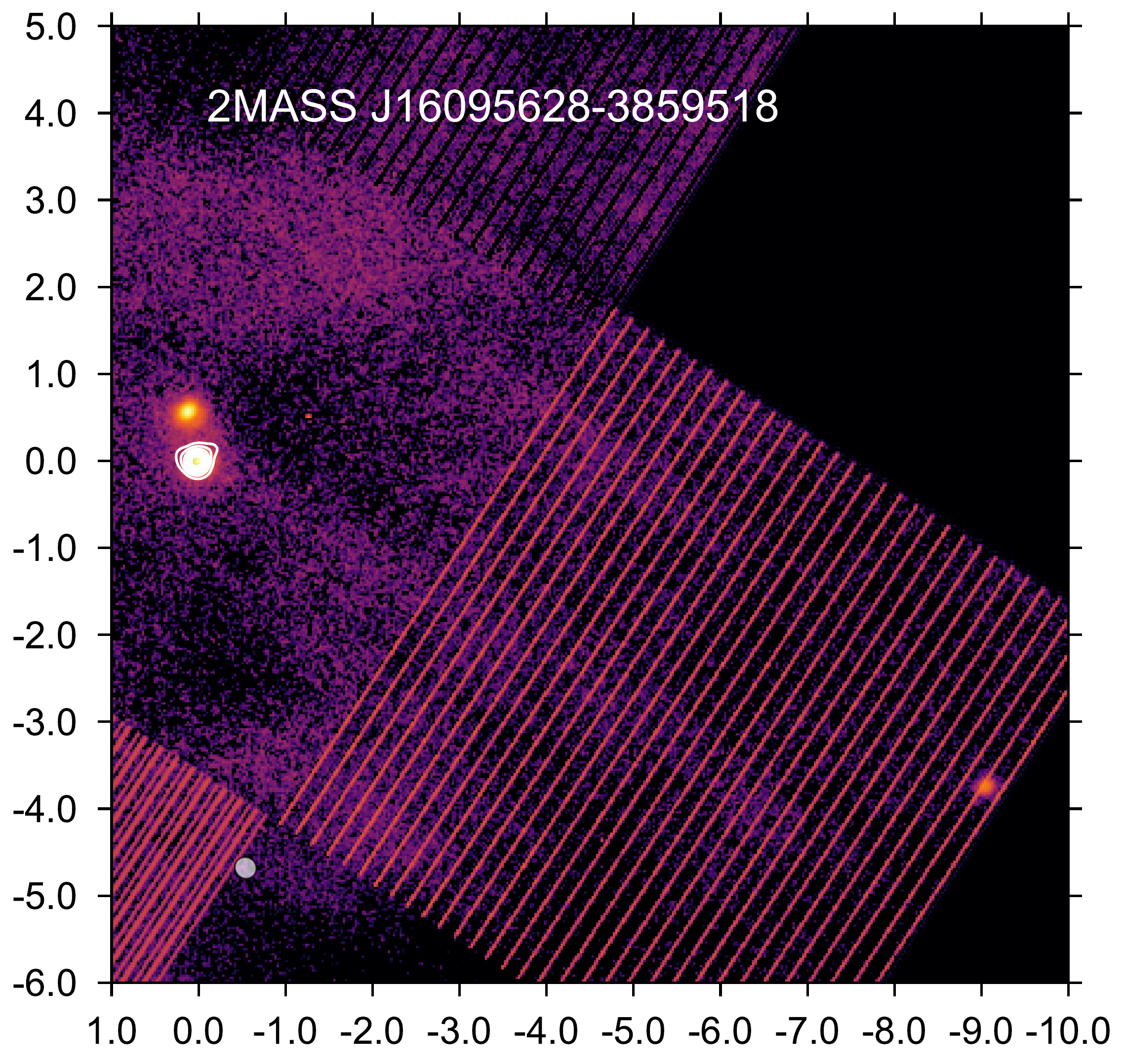} \hfill
\includegraphics[height=0.3\textwidth]{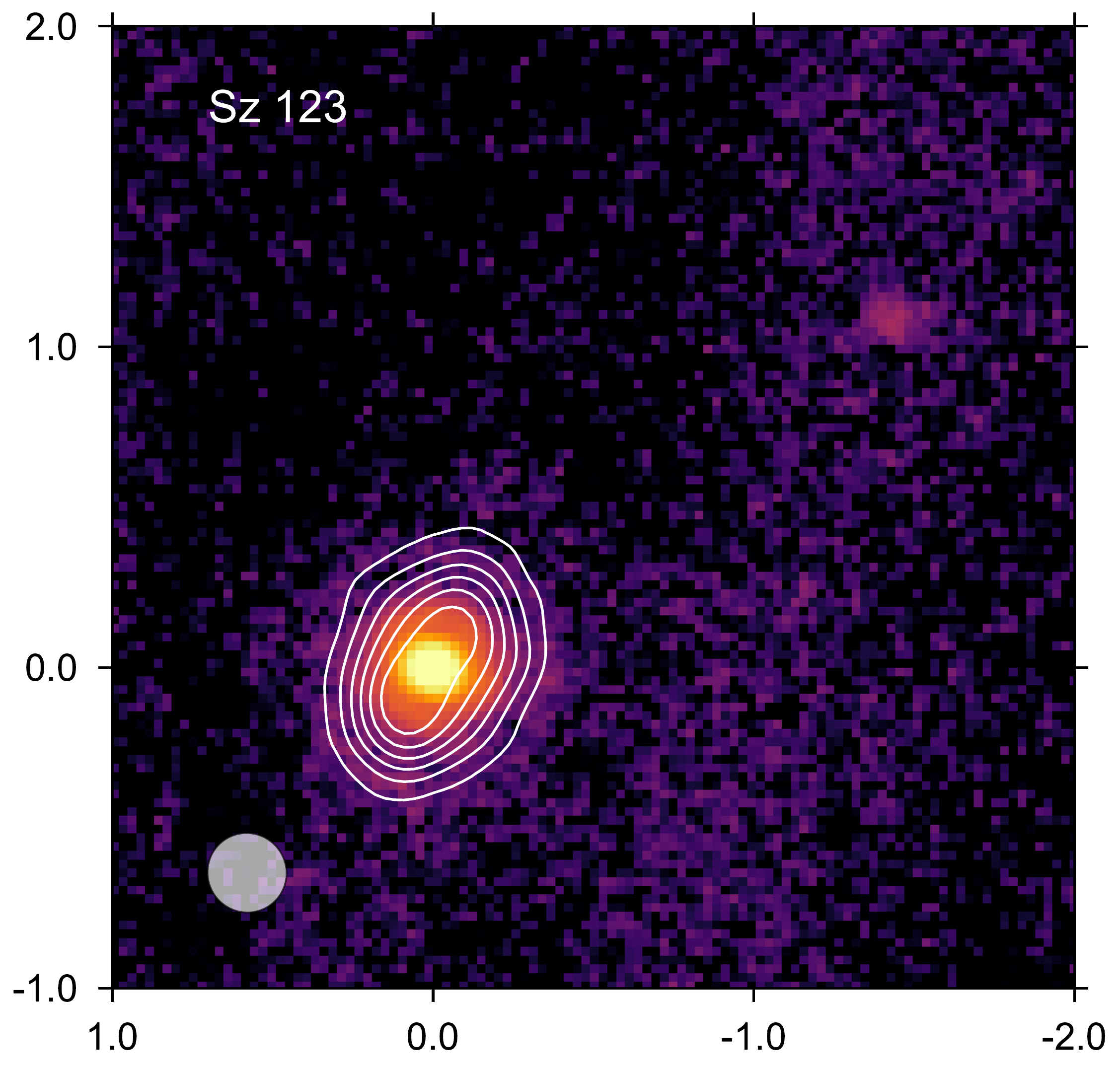}  \\
\includegraphics[height=0.3\textwidth]{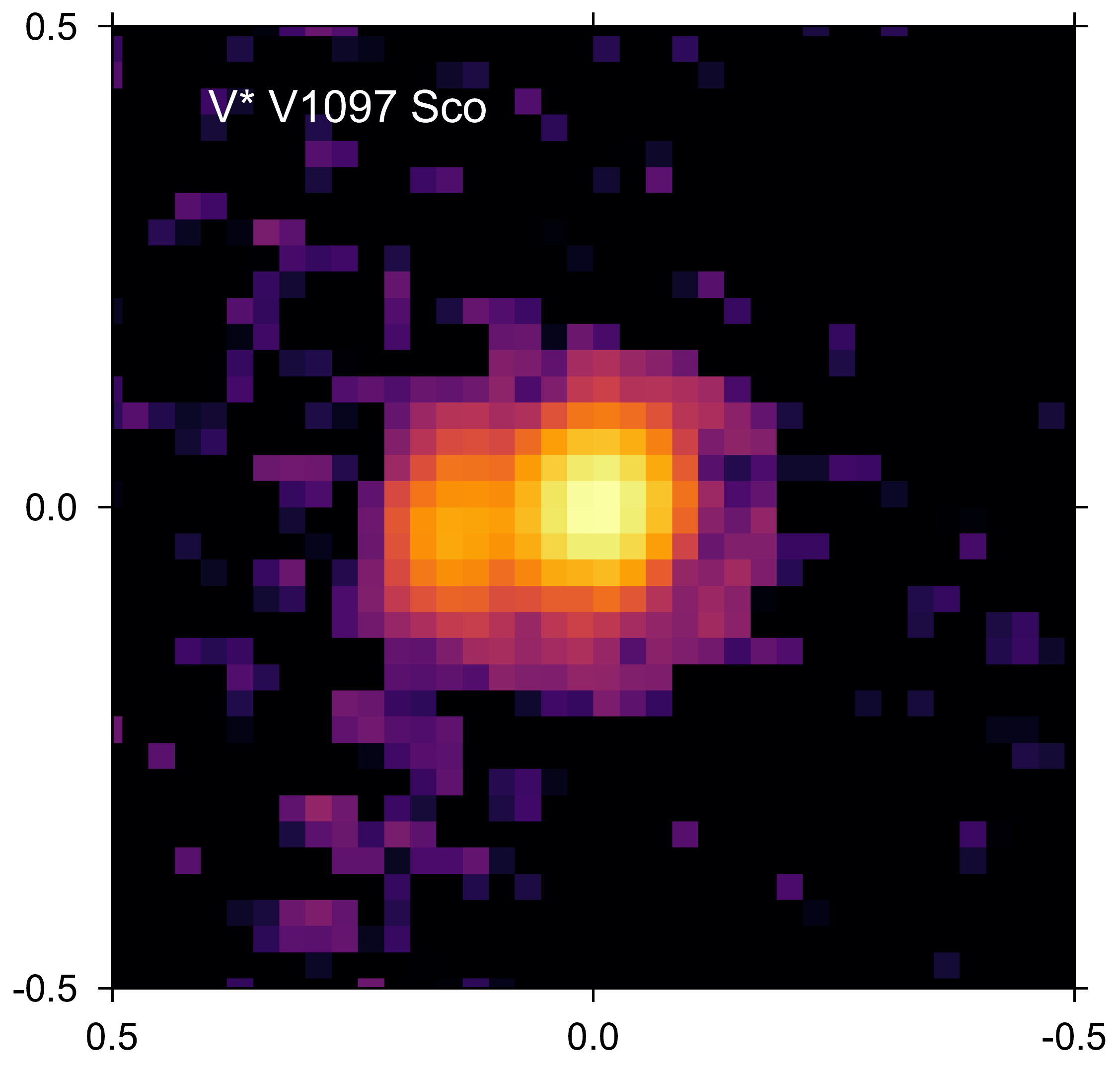} 
\caption{Same as Figure~\ref{f:mos}. 
}
\label{f:mos2}
\end{figure*}

\begin{figure}
    \centering\includegraphics[width=\columnwidth]{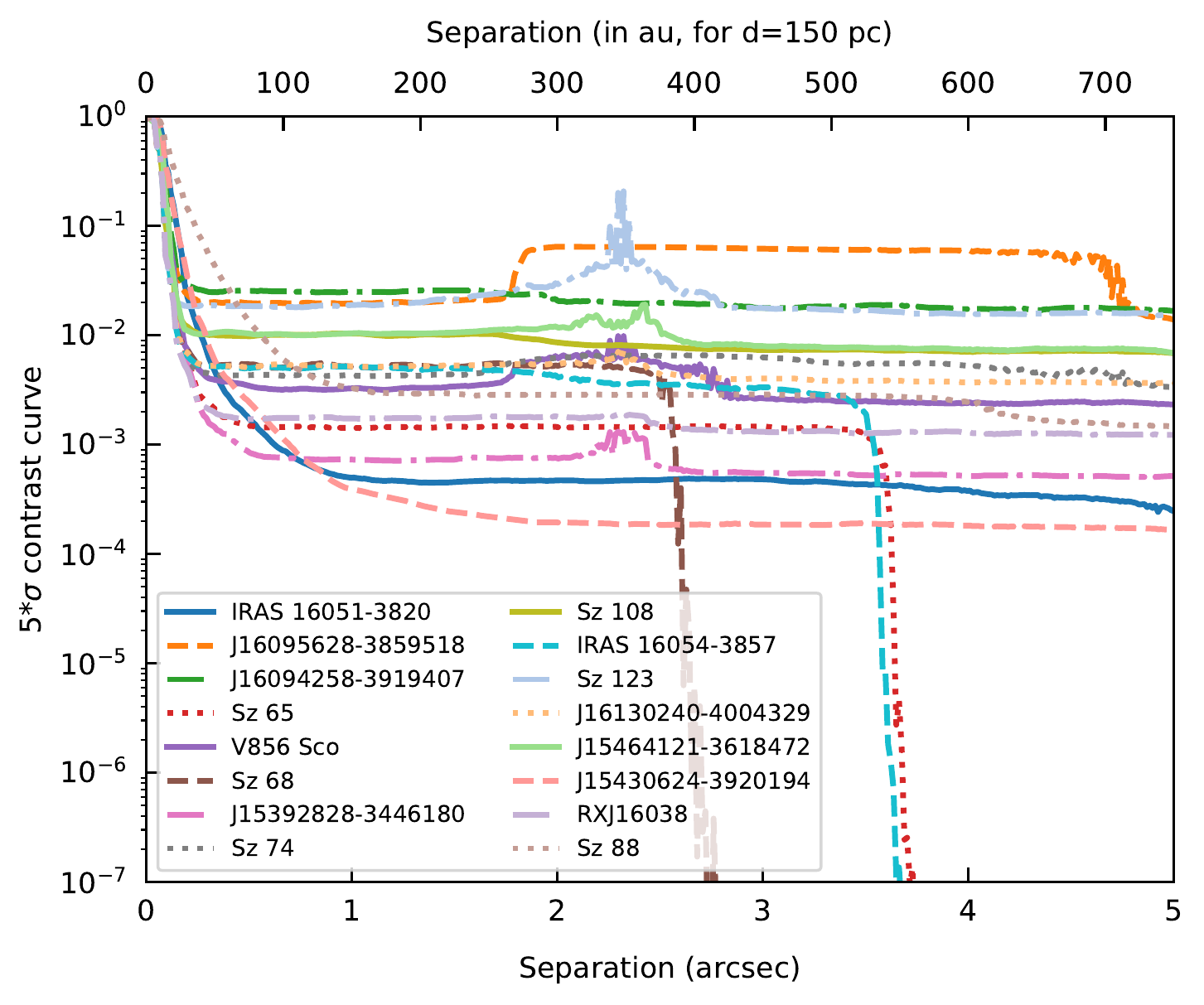}
    \caption{{ Contrast curves extracted from images shown in Figure~\ref{f:mos} and Figure~\ref{f:mos2}, calculated from azimuthally-averaged profiles centered at the primary's location. The curves flatten at distances greater than 5\arcsec. We note that the structures displayed by some curves beyond 2\arcsec are due to detector artifacts that affect the averaged profile, or due to reaching the edge of the detector.  Labels are provided for all curves. }}
\label{f:contr}
\end{figure}







\begin{table*}
  \begin{minipage}{1\textwidth}
\caption{{Properties of the companions detected by the NIR-Lupus survey. Note that the RA and DEC coordinates are reported for the primary star. Objects marked with a $\blacktriangledown$ are in the planetary mass regime, in the range 8-12 \MJup. The distance marked with $^{*}$ is not derived from \emph{Gaia} DR2 \citep{Bailer_Jones_2018} as the others, but assumed to be 150 pc, the mean distance for the Lupus star forming region. { The uncertainty on the separation is 2 au for all the targets, and on the angle is 2 degrees, as explained in Section~\ref{s:obsnaco}.}  }} 
\label{t:bin}
\centering
{\footnotesize
\begin{tabular}{lccccccccccc}
\hline
\hline
Name & RA & DELoC  & Class & dist & Sep  & PA  & NIR & mm & Cat.\footnote{RES: resolved, UN: unresolved, ND: non-detected discs with ALMA. See Table~\ref{t:cat} and Section~\ref{s:res} for more details.} & $P_{\rm bkg}$\footnote{Probability of being a background object. We indicate known bound companions with a dash sign.} & New or\\
 &   &    & & (pc) &  (au) & (deg) & $f_2/f_1$ & $f_2/f_1$ & & (\%) & Literature\\
\hline
Sz65B (Sz66) &		234.8656	&	-34.7715	&II &	155	&	983	&	97	&	0.46	&0.21 & Res	& 0.1 & L\\
	2MASS~J15430624-3920194B &	235.7760	&	-39.3387	&III &	165	&	{ 24}	&	-16	&	0.99	&	-& ND	& $< 10^{-4}$  & N \\
Sz68B\footnote{Archive NB\_1.64 filter data}		&	236.3027	&	-34.2916	&II &	154	&	17	&	-90	&	0.10	&	- & UN	& -& N \\
Sz68C$\color{blue}^c$		&	236.3027	&	-34.2916	&II &	154	&	433	&	-63	&	0.07	&	0.05 & Res	& - & L\\
	2MASS~J15464121-3618472B	&	236.6717	&	-36.3132	&III &	149	&	172	&	88	&	0.13	&	-& ND	& $< 10^{-4}$ & N\\
Sz74B		&	237.0217 &	-35.2648	&II &	150$^{*}$	&	49	&	-5	&	0.33	&	0.03  & UN	& $< 10^{-4}$ & L\\
Sz81B\footnote{Archive K-band data}		&	238.9594	&	-38.0261	&II &	160	&	309	&	19	&	0.62	&	0.34 & Res	& $< 10^{-4}$ & L\\
RXJ16038B$\color{blue}^d$	&	240.9687	&	-39.6504	&III &	130	&	41	&	-97	&	0.16	&	-	& ND & $< 10^{-4}$& N\\
Sz88B		&	241.7524	&	-39.0389	&II &	158	&	245	&	34	&	0.1	&	- & ND	& 0.1& L\\
IRAS 16051-3820B$\color{blue}^d$		&	242.1279	&	-38.4742	&II &	156	&	787	&	-31	&	0.2	&	-	& ND & 0.2& N\\
V856ScoB		&	242.1428	&	-39.1052	&II &	161	&	231	&	111	&	0.02	&	0.36 & Res	& $< 10^{-4}$ & N \\
Sz108B		&	242.1786	&	-39.1042	&II &	169	&	683	&	23	&	0.31	&	-& N.B.\footnote{Not bound, see Table~\ref{t:gaia}.}	& 0.2 & L\\
IRAS 16054-3857B		&	242.2058	&	-39.0944	&II &	159	&	69	&	8	&	0.77	&	-& ND	& $< 10^{-4}$ & N\\
	2MASS~J16094258-3919407B	&	242.4274	&	-39.3280	&III &	147	&	245	&	26	&	0.6	&	-& { N.B.$\color{blue}^e$}	& $< 10^{-4}$ & N\\
	2MASS~J16095628-3859518B	&	242.4845	&	-38.9978	&II &	157	&	90	&	12	&	0.93	&	-& ND	& $< 10^{-4}$& N\\
	$\blacktriangledown$2MASS~J16095628-3859518C	&	242.4845	&	-38.9978	&II &	156	&	1529	&	-112	&	0.14	&	-& ND	& 16.4& N\\
$\blacktriangledown$Sz123C		&	242.7149	&	-38.8873	&II &	162	&	292	&	-53	&	0.01	&	-& { N.B.$\color{blue}^e$}	& 1.0& N\\
V*V1097 ScoB	&	243.2601	&	-40.0758	&III &	139	&	21	&	100	&	0.35	&	-& ND	& $< 10^{-4}$ & N\\

\hline
\end{tabular}
}
  \end{minipage}
\end{table*}

\begin{table*}
\caption{{Properties of the companions in Lupus detected by ALMA but not observed or detected in the NIR. These candidate secondaries were assumed as contaminants in our analysis.}} 
\label{t:bin_alma}
\centering
\begin{tabular}{lccccc}
\hline
\hline
Name    & RA & DEC  & Sep (au) & PA (deg) & mm flux ratio \\
\hline
 Sz88B?     &     241.7524	&	-39.0389 &  54 & 213  & - \\ 
J16070384-3911113B? &  241.7649 &  -39.1867 & 568 & 268  & 0.39 \\
J16073773-3921388B? & 241.9065  & -39.3609 & 344 & 267  & 0.87  \\

\hline
\end{tabular}
\end{table*}

\section{Results}
\label{s:res}

We identify a total of { 16} multiple systems: { 3} triple systems and { 13} binaries. { Three of these detected companions} (Sz~108B, { Sz~116 or 2MASS J16094258-3919407B, and Sz 123C) are at different distances than their primaries, thus we consider them not physically bound (see below and Table~\ref{t:gaia})}. Some of these systems were already known as multiples from the literature (see Section~\ref{s:sample}). The NIR images, over-plotted with the ALMA contours, are shown in Figures~\ref{f:mos} and \ref{f:mos2}. Four systems are resolved with ALMA, two close binaries are unresolved with ALMA, and { 9} systems do not show millimetre flux in at least one of the components. Each multiple system is labelled as resolved (Res), unresolved (UN), not detected (ND), { or not bound (N.B.)} in Table~\ref{t:bin}. A summary of these 3 categories of multiple system is presented in Table~\ref{t:cat}, together with the number of single stars. For each multiple system we measure the separation and the position angle of the companions, along with the ratio of NIR flux. The millimetre flux ratio is also reported from \cite{Ansdell2018}. Most of the systems have \emph{Gaia} Data Release 2 (DR2) parallax measurements \citep{Bailer_Jones_2018}, which we report in Table~\ref{t:bin} with all the other information mentioned above { for the primaries}. { Systems for which \emph{Gaia} detected the secondary component are listed in Table~\ref{t:gaia}; the systems that present discrepancies in the distances of the components are assumed as not physically bound.}

\begin{table}
\caption{ {\emph{Gaia} distances for the detected systems. Objects assumed as not physically bound are marked with a *. }} 
\label{t:gaia}
\centering
\begin{tabular}{lc}
\hline
\hline
Name    & \emph{Gaia} distance (pc) \\
\hline
Sz~65 & 155.3 \,$\pm$\, 1.2 \\
Sz~66 &  157.3 \,$\pm$\, 2.1 \\
\hline
Sz~68A &  154.2 \,$\pm$\, 1.4 \\
Sz~68B & 154.9 \,$\pm$\, 3.0 \\
\hline
Sz~81A &  159.9 \,$\pm$\, 1.6 \\
Sz~81B &   160.0 \,$\pm$\, 2.1\\
\hline
Sz~88A &  158.4 \,$\pm$\, 1.2\\
Sz~88B &  159.0 \,$\pm$\, 2.7\\
\hline
* Sz~108A & 151.3 \,$\pm$\, 1.2 \\
* Sz~108B &  169.0 \,$\pm$\, 3.4 \\
\hline
* 2MASS J16094258-3919407 & 146.5 \,$\pm$\, 0.8 \\
* 2MASS J16094258-3919407B (Sz 116) &  154.3 \,$\pm$\, 1.8 \\
\hline
* Sz~123AB &  162.8 \,$\pm$\, 1.2 \\
* Sz~123C  &  1497.0 \,$\pm$\, 322.3  \\
\hline
\end{tabular}
\end{table}

For { the other companion candidates that remain unconfirmed as bound}, we estimate their probability of being background objects. We first consider a homogeneous spatial Poisson point process to calculate the probability $\mathcal{P}$($n$=$0|\lambda, B$) of not having any background object that is of equal or greater brightness to that of the companion candidate in an area $B$ for a number density $\lambda$. For each companion candidate, we set $B$ to the area of a circle with a radius equal to the candidate separation, and estimate the number density of background stars that are of equal or greater brightness in that area of the sky using the TRILEGAL model of the Galaxy \citep{Girardi2012}. We then consider the probability of having at least one background star by chance $\mathcal{P}$($n>0|\lambda, B$) = 1 - $\mathcal{P}$($n$=$0|\lambda, B$) to be a reasonable proxy for the probability of the candidate being a background object, as in \citet{Reggiani2018} and \cite{Ubeira-Gabellini2020}. The calculated probabilities are reported in the last column of Table \ref{t:bin}. For all stellar candidates, the probabilities are approximately null owing to their large fluxes. 

Sz~88 presents an elongated shape in the millimetre, which may be attributed to an edge-on disc or a close companion. The close millimetre blob is only marginally detected in the ALMA data at the $\sim$ 3$\sigma$ level, and no signal is detected at the same position in the NIR. The nature of the close millimetre blob should therefore be further investigated with deeper high-contrast and/or high-resolution observations. Sz 123AB is composed of two close M-type stars \citep{alcala2014} and we detect a third faint object in the North West. { This faint companion is detected by \emph{Gaia} at a distance of 1497.0 \,$\pm$\, 322.3 (Table~\ref{t:gaia}), thus we assume it is a background object.} Note that the central binary appears as a single star in the NIR, while in the millimetre the PSF is elongated, showing what seems to be a circumbinary disc. Another faint companion is detected as a tertiary component in the J16095628-3859518 system. From the AMES-COND evolutionary models, { following the same procedure used in \cite{Zurlo2020},} the object would have an estimated mass of 12 \MJup. The candidate planet is located at $\sim$ 1500 au from the binary system. 
We estimate the probability of this object being a background star to be $\sim$17\%, hence a second-epoch observation is needed to rule out this possibility.

\begin{table*}
\caption{{Summary of the four categories in which the population of NIR-Lupus is divided. Note that there are { 2} triple systems, so the total number of systems are 105. }} 
\label{t:cat}
\centering
\begin{tabular}{lcl}
\hline
\hline
Category & N. of objects   & Description  \\
\hline
Single objects    &  { 92}	& Single class II and class III objects in Lupus	\\
Resolved discs (Res)  & 4	&  Resolved disc companions in the mm 	\\
Unresolved multiple discs (UN)          & 2	&Binary systems where the components are unresolved in the mm 	\\
ND of component(s) (ND) & { 9}	&	Only the primary is detected in ALMA, or none of the components is detected	\\
\hline
\end{tabular}
\end{table*}

\begin{figure}
    \includegraphics[width=\columnwidth]{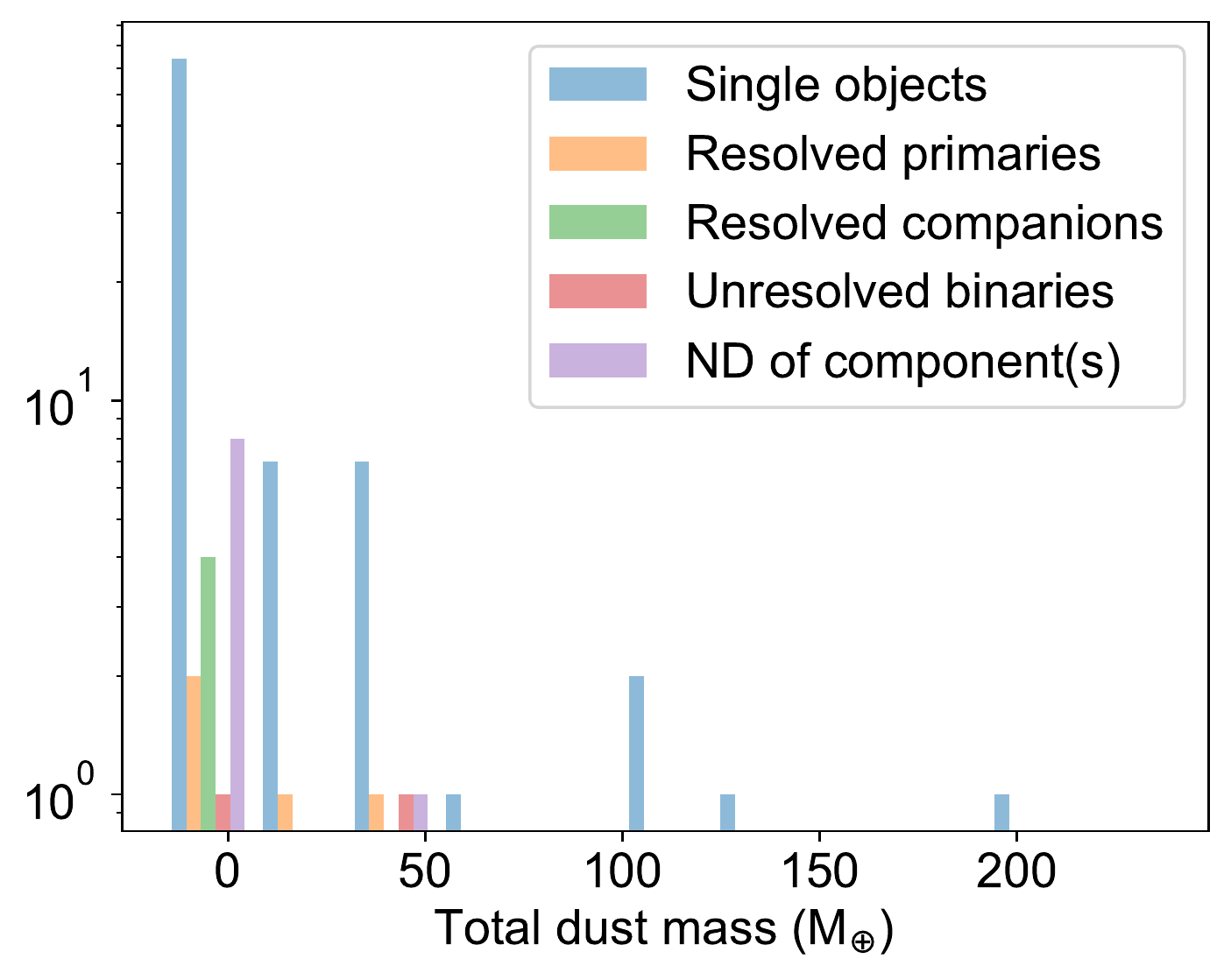}
    \caption{Histogram of the total mass of the dust of the discs measured in the ALMA data.  }
    \label{f:histo}
\end{figure}

\begin{figure*}
    \includegraphics[width=\columnwidth]{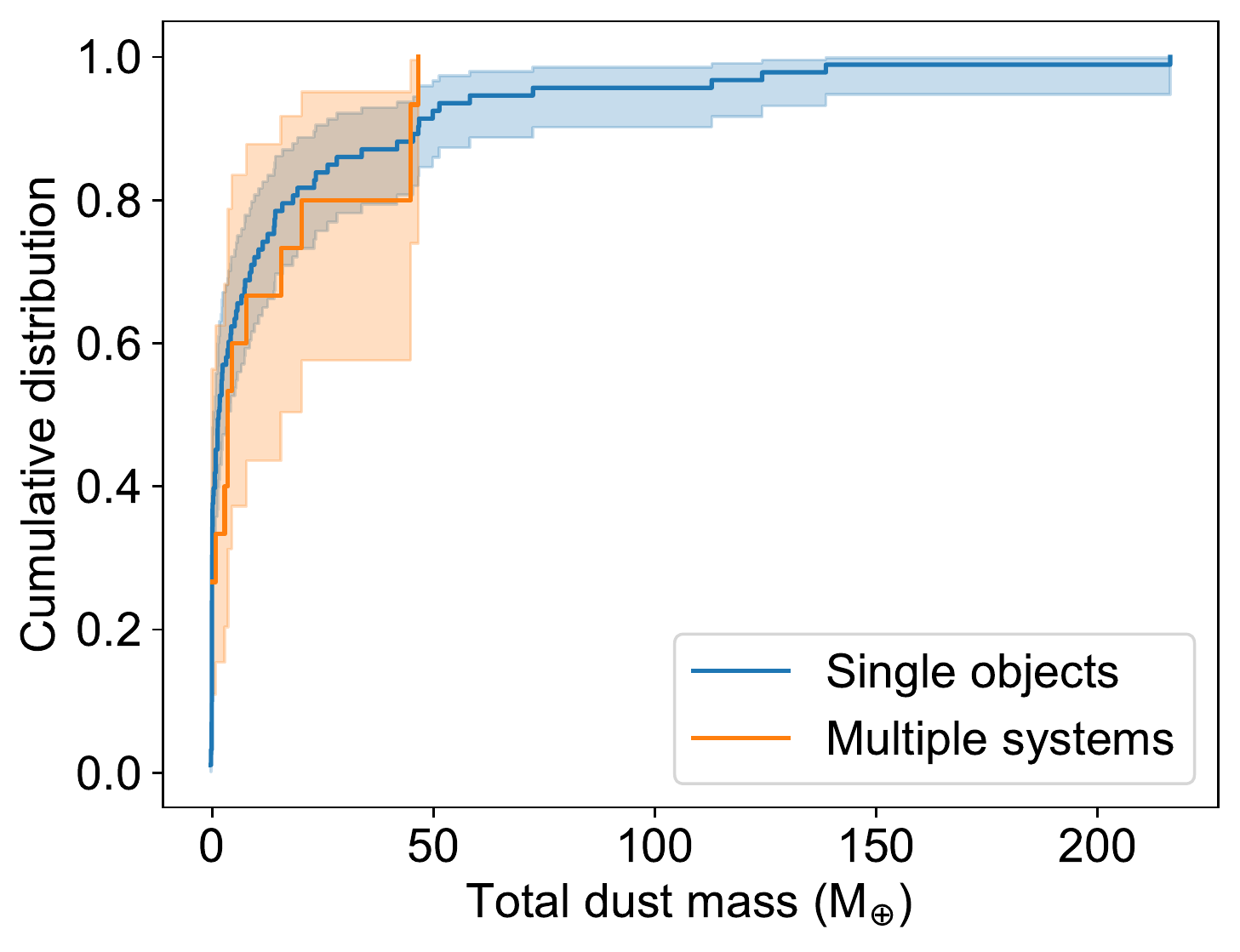}\hfill
    \includegraphics[width=\columnwidth]{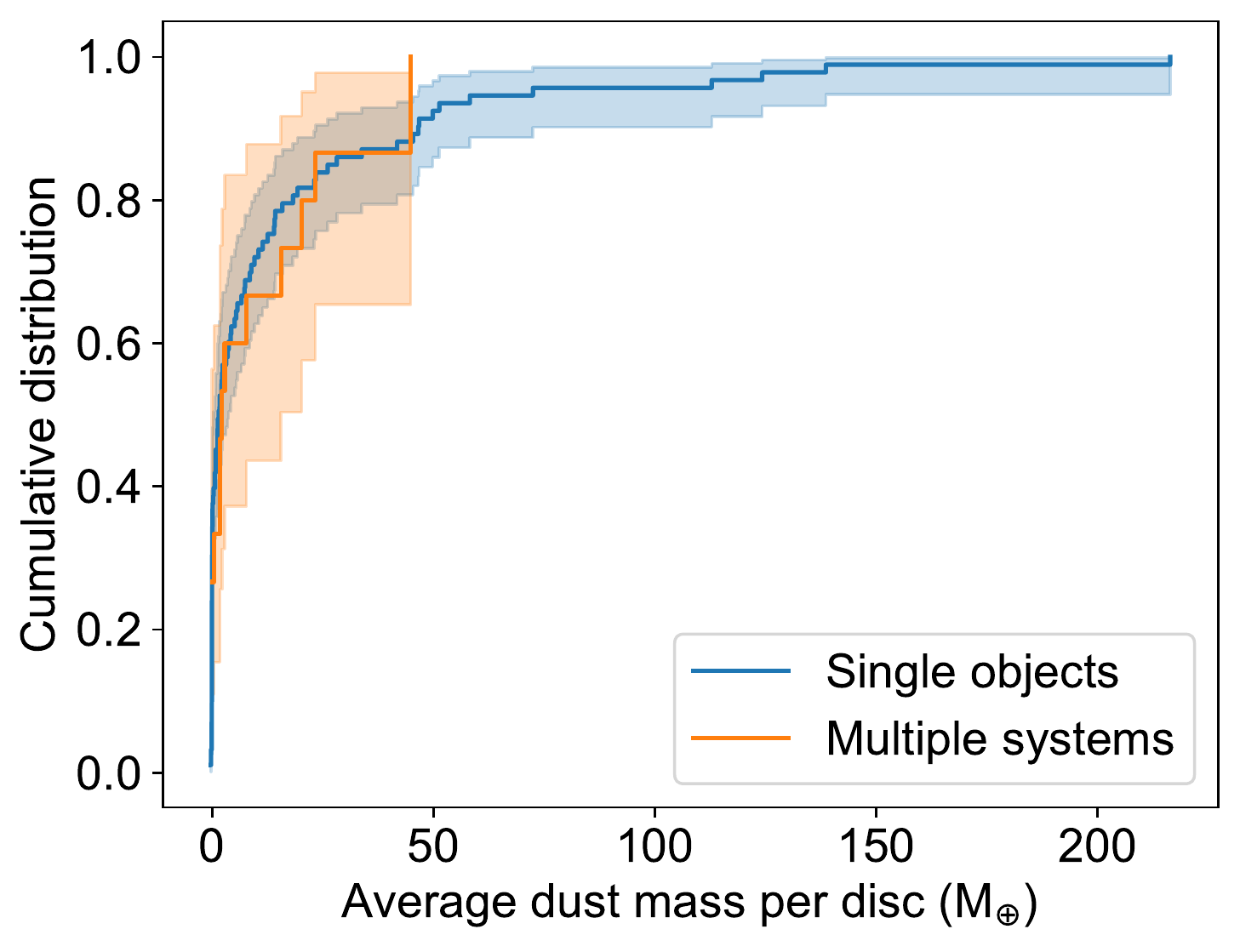}
    \caption{ Cumulative distribution of the total mass of the dust ({\it left}) and of the average mass of the dust around each star ({\it right}) in the systems as measured in the ALMA data, single vs multiple systems are shown.}
\label{f:cum_single}
\end{figure*}

\begin{figure}
    \includegraphics[width=\columnwidth]{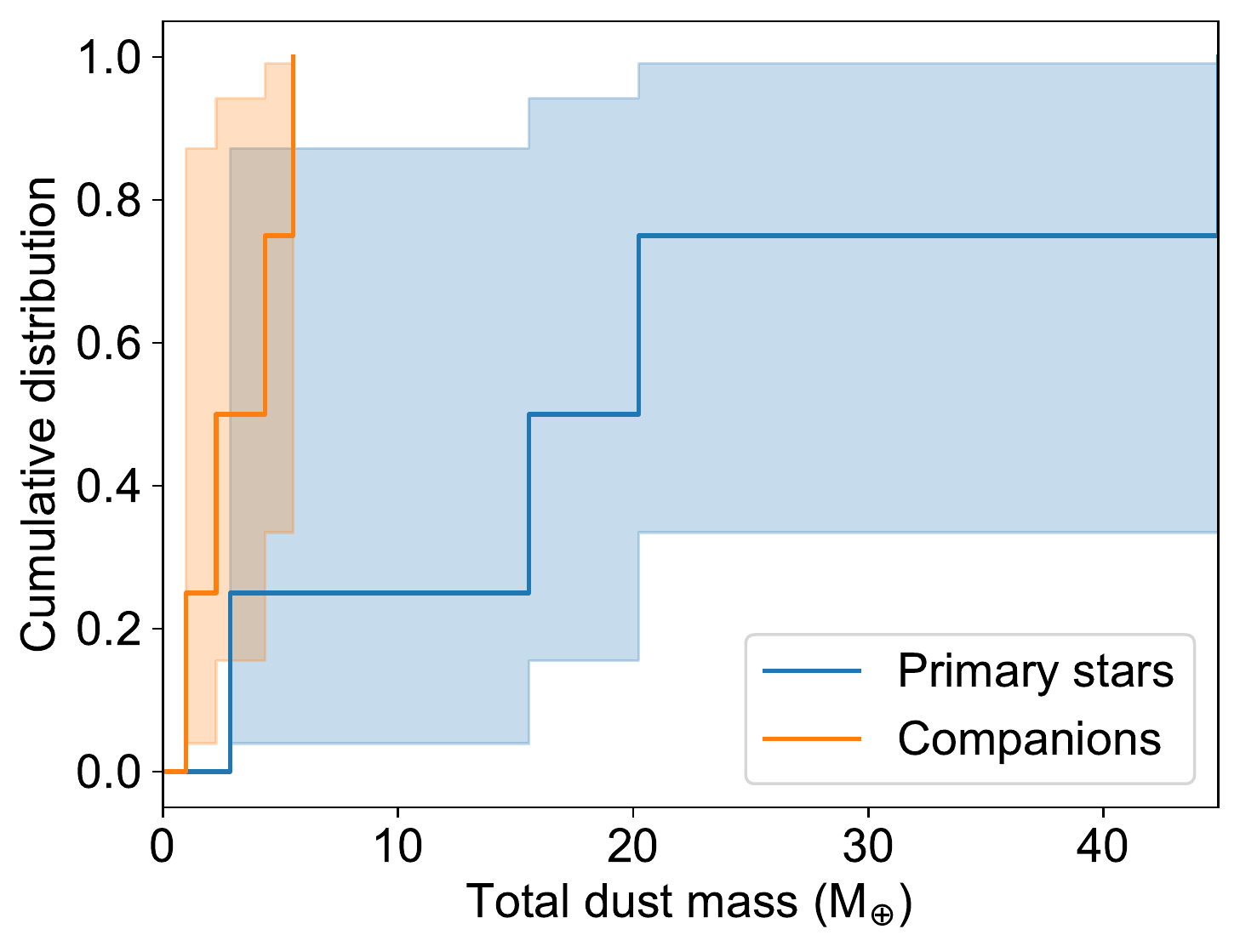}
    \caption{Cumulative distribution of the total mass of the dust in the primaries vs the companions as measured in the ALMA data.}
    \label{f:cum_a}
\end{figure}

\begin{figure}
    \includegraphics[width=\columnwidth]{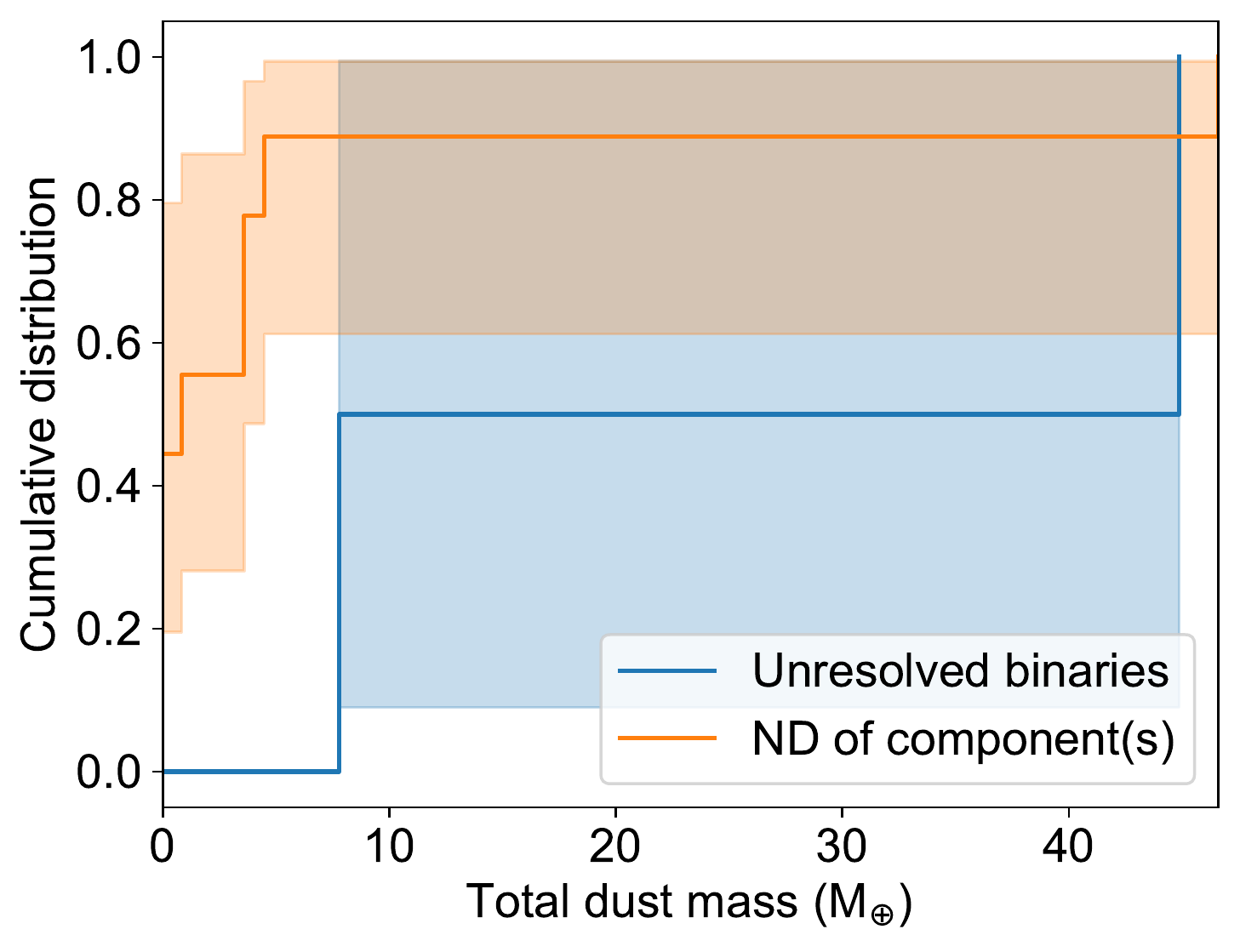}
    \caption{Cumulative distribution of the total mass of the dust in the systems where the tight binaries are not resolved by ALMA vs the ones where the binaries are resolvable in the mm, but only the primary are detected.}
\label{f:cum_unre}
\end{figure}

\begin{figure}
    \includegraphics[width=\columnwidth]{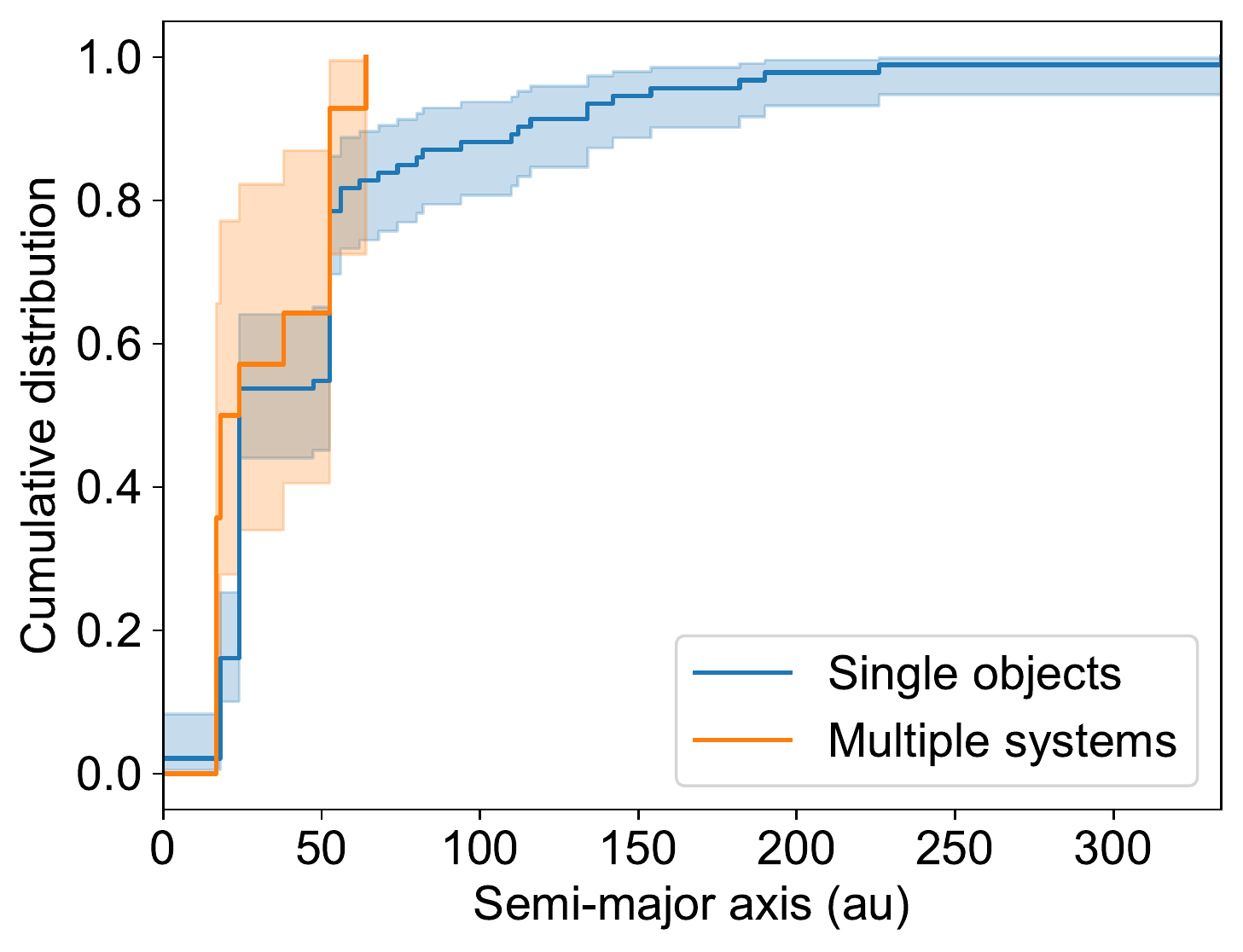}
    \caption{Cumulative distribution of the projected semi-major axis of the discs as measured in the ALMA data.}
    \label{f:cum_semi}
\end{figure}

\begin{figure*}
    \includegraphics[width=\columnwidth]{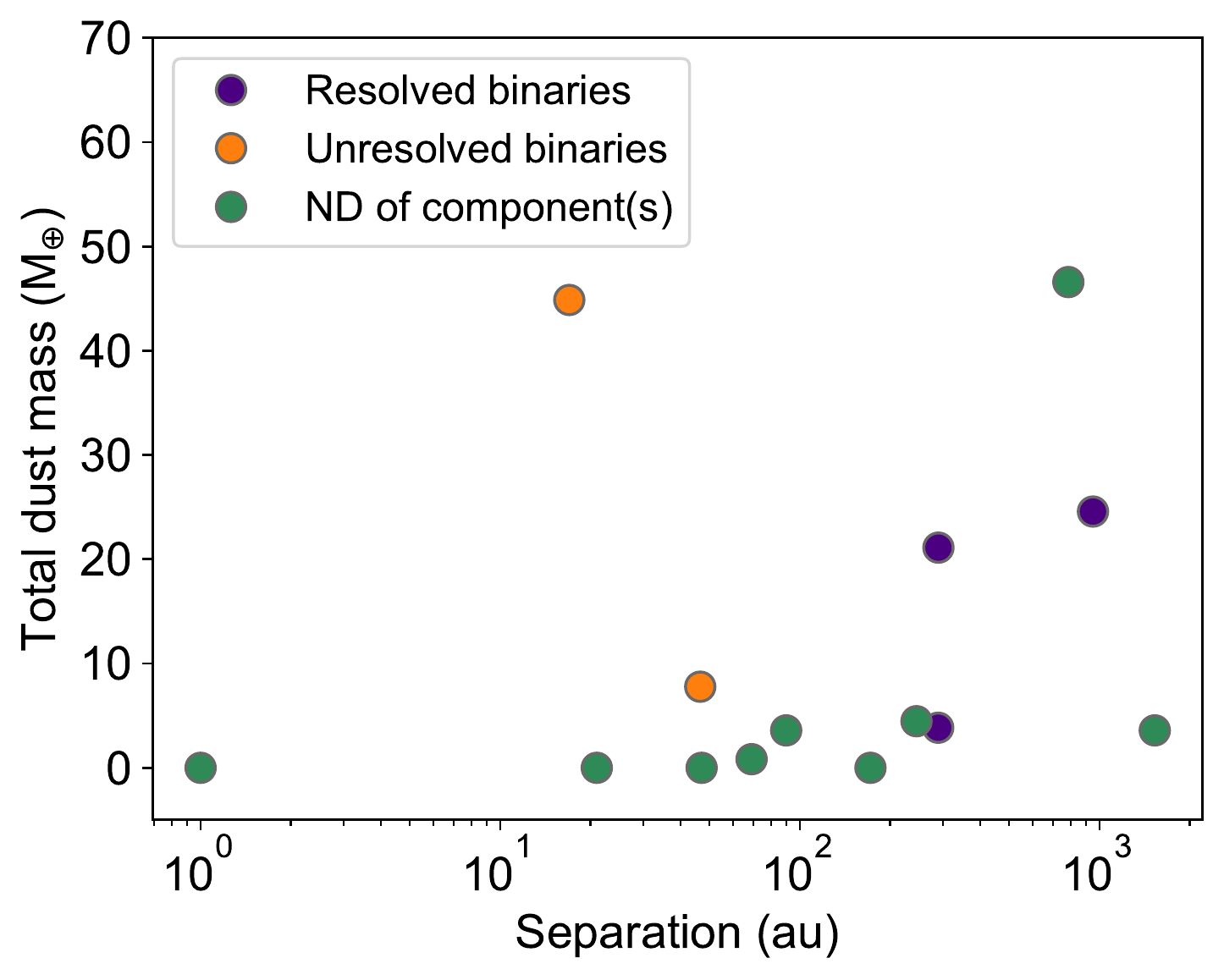} \hfill
    \includegraphics[width=\columnwidth]{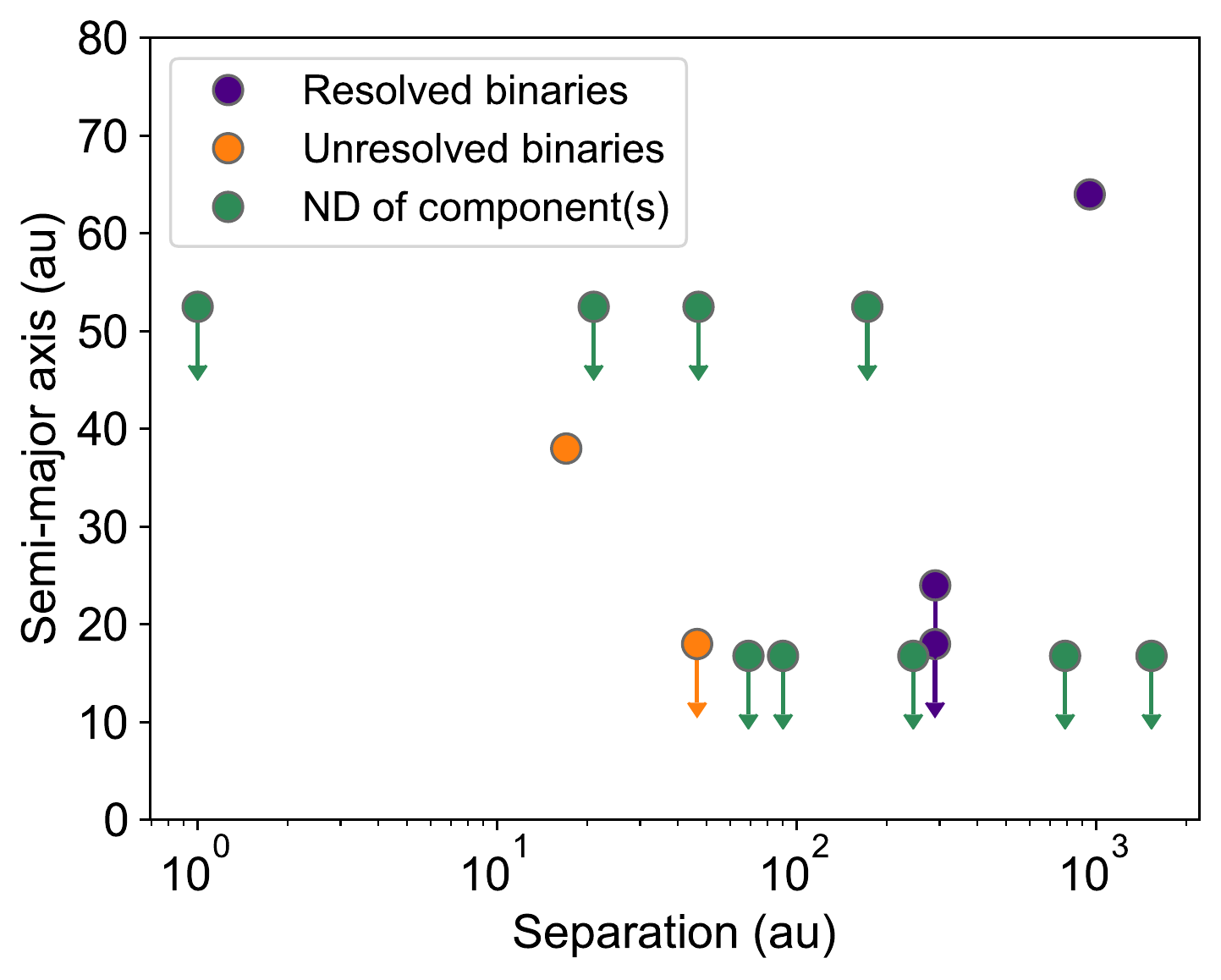}
    \caption{Total mass of the system in dust ({\it left}) and semi-major axis of the discs as measured in the ALMA data ({\it right}) versus projected separation in between each component of the multiple system. Unresolved discs are indicated with arrows. }
    \label{f:a_sep}
\end{figure*}

\begin{figure}
    \includegraphics[width=\columnwidth]{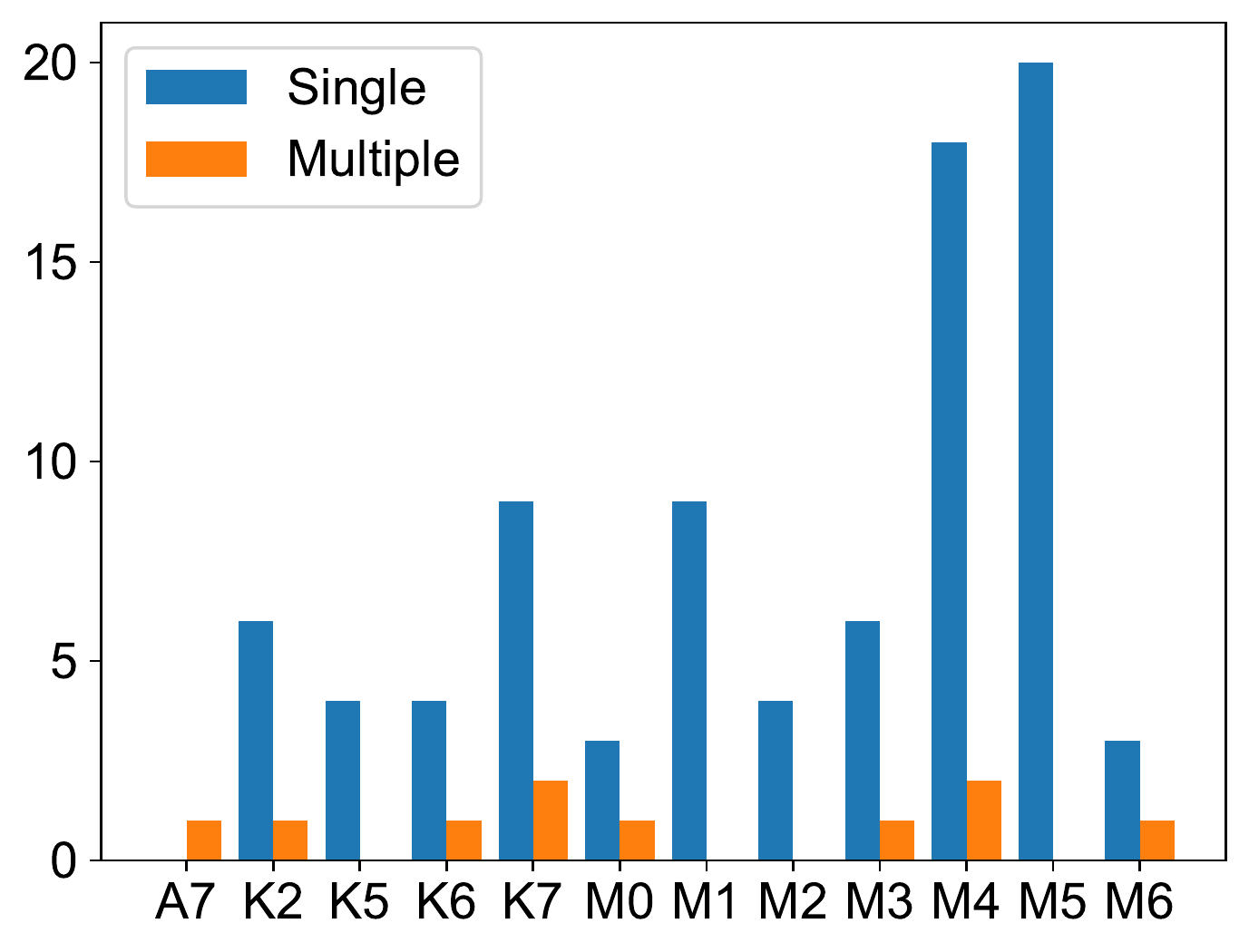}
    \caption{Histogram of the spectral type of the single stars vs multiple systems. }
    \label{f:histo_sp}
\end{figure}

As for NIR-ODISEA for the Ophiuchus molecular cloud, we compare the masses of the Lupus discs in the different multiplicity categories. The dust mass for each disc is taken from { the last column of Table 1 in} \cite{Ansdell2018} for the Class II objects, and from { Table 1 of \citet{2020MNRAS.tmp.3151L}} for the Class III objects. { Only four companions are detected in the millimetre: Sz~68, Sz~81A, Sz~65, and V856~Sco. The values of their masses are obtained from the millimetre fluxes listed in Table 4 of \cite{Ansdell2018}, according to the same procedure described by the authors for the primary components}. Figure~\ref{f:histo} shows a histogram of the disc masses in Lupus, color-coded by multiplicity category. In Figure~\ref{f:cum_single}, we show cumulative distribution functions of disc mass, accounting for non-detections using the python package {\sc lifelines} \citep{2019JOSS....4.1317D}, a Kaplan-Meier (KM) product estimator, as in \cite{Zurlo2020}; for clarity, we present separately the distributions for the dust mass per system (left) and per component (right). We also compare the dust mass cumulative distributions for the primary vs companion components (Figure~\ref{f:cum_a}) and with unresolved and undetected components (Figure~\ref{f:cum_unre}). The cumulative distribution of the disc sizes for single and multiple objects, expressed in semi-major axis of the disc, is displayed in Figure~\ref{f:cum_semi}. The disc size for the Class II objects is used when available in \cite{Ansdell2018} { (Table 3)}, otherwise we assume the disc is unresolved. Concerning the Class III objects, 3 discs were unresolved, and only one was marginally resolved \citep[0\farcs7,][]{2020MNRAS.tmp.3151L}. The dust radii are measured from the 1.33 mm continuum images using a curve-of-growth method, in which successively larger photometric elliptical apertures are applied until the measured flux is 90\% of the total flux. For the detected multiples we show, respectively, the total mass (Figure~\ref{f:a_sep}, left panel) and the semi-major axis (Figure~\ref{f:a_sep}, right panel) of the primary disc vs the separation of the companion, to display any dependence of the mass and size of the discs with the separation of the components in the system. Finally, in Figure~\ref{f:histo_sp} we show the distribution of the spectral type of the primary stars in single and multiple systems.  

\section{Discussion}
\label{s:discussion}

Even though the NIR-Lupus sample (105 objects, { 20 discs are not observable with NACO}) is smaller than that of Ophiuchus \citep[236 objects, presented in][]{Zurlo2020}, the same trends are found. 
The fraction of { visual} multiple systems { (with separations of 20--4800 au)} are similar in both samples, with { 12$\%$ (13/105)} in Lupus and 18$\%$ (43/236) in Ophiuchus. These fractions are significantly lower than { in} other 
star-forming regions where the multiplicity censuses are more complete. The Taurus region, for example, shows a multiplicity frequency of $\sim$70\%, when including spectroscopic binaries \citep{2011ApJ...731....8K}. Therefore, the multiplicity fraction in Lupus and Ophiuchus is likely to increase { once} more spectroscopic and/or very tight binaries are identified.

\begin{figure}
\begin{center}
\includegraphics[width=\columnwidth]{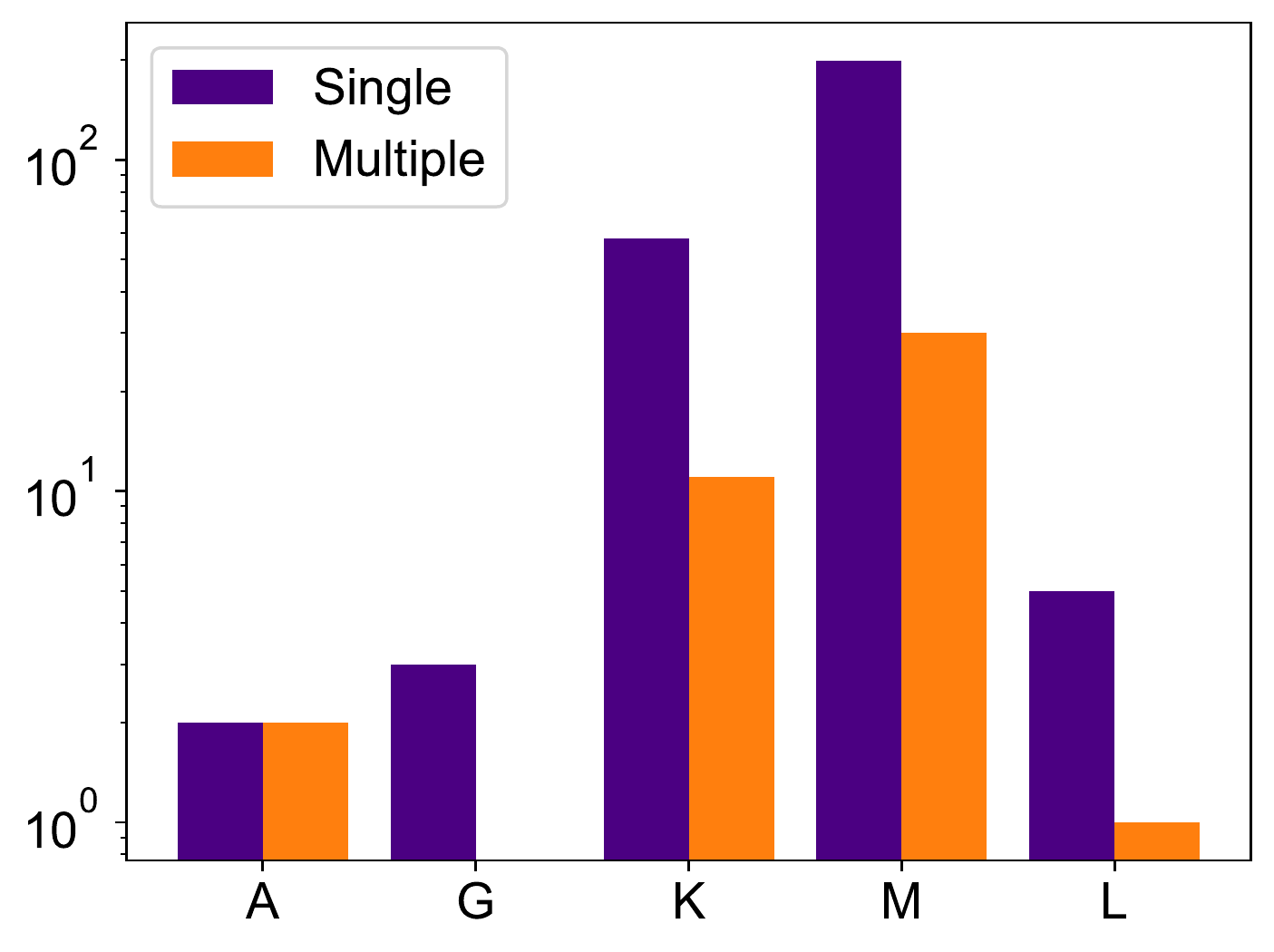}
\caption{Histogram of the spectral type of the single stars vs multiple systems of all the primary stars of Lupus and Ophiuchus. The sample is dominated by M-type stars, but multiple systems have been identified across the entire spectral range.}
\label{f:histo_sp_com}
\end{center}
\end{figure}

For the Lupus region, single and multiple systems are found to be homogeneously distributed by spectral type, as shown in Figure~\ref{f:histo_sp}, but the fraction of multiple systems is lower for late-type objects, as expected. In Figure~\ref{f:histo_sp_com} we show the histograms that include also the Ophiuchus sample, which follows the same trend.  Stellar multiplicity surveys show that the frequency of multiple systems strongly depends on the mass of the primary \citep[e.g.,][]{1991A&A...248..485D, 1992ApJ...396..178F,2003ApJ...586..512B}. Indeed, solar-type stars have a high probability of having companions \citep[$\sim$50\%,][]{2010ApJS..190....1R}, while very low mass objects (T5-Y0 spectral types) show a significantly lower fraction of 8\% { \citep{2013ARA&A..51..269D}}.

\begin{figure*}
\begin{center}
\includegraphics[width=\columnwidth]{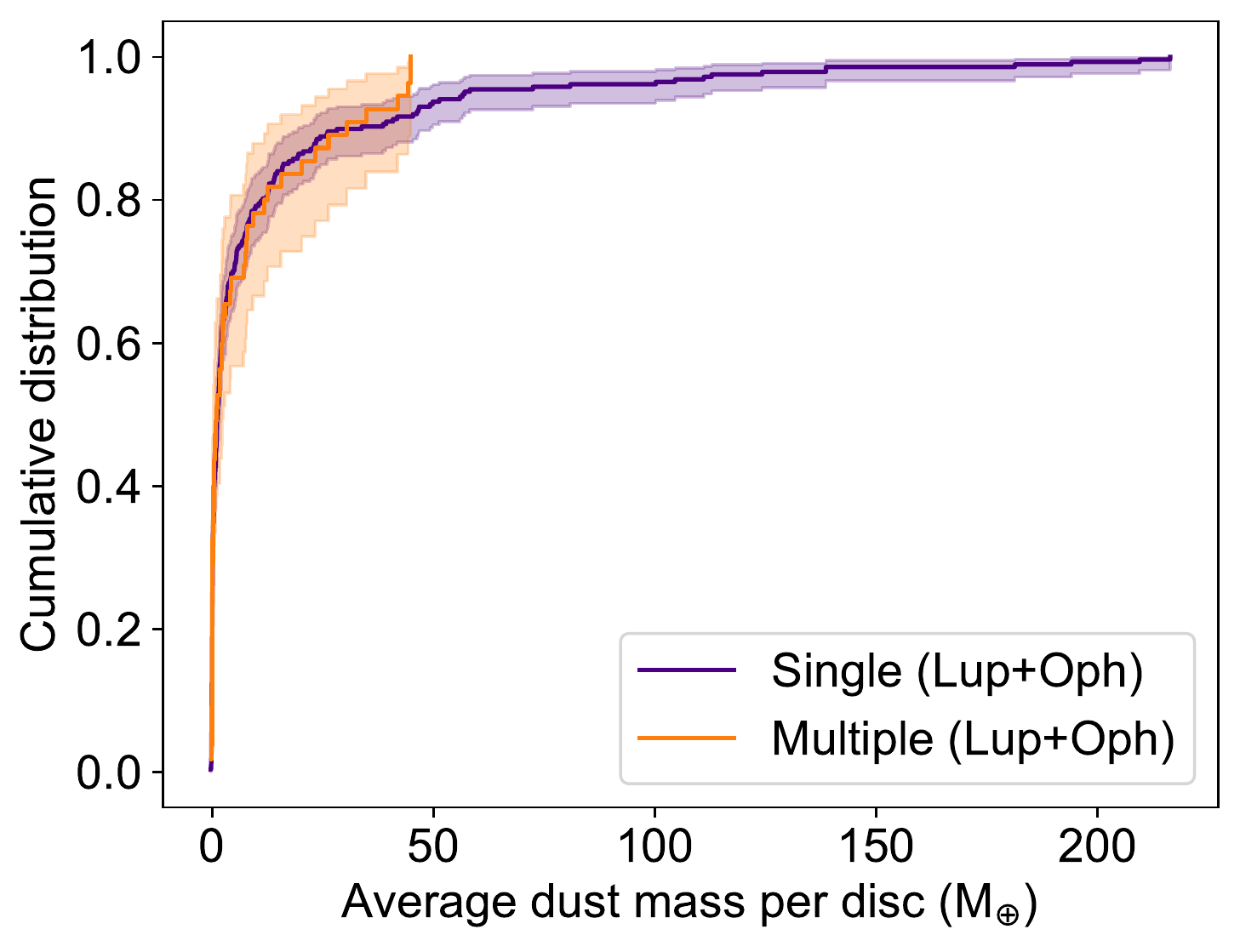} \hfill
\includegraphics[width=\columnwidth]{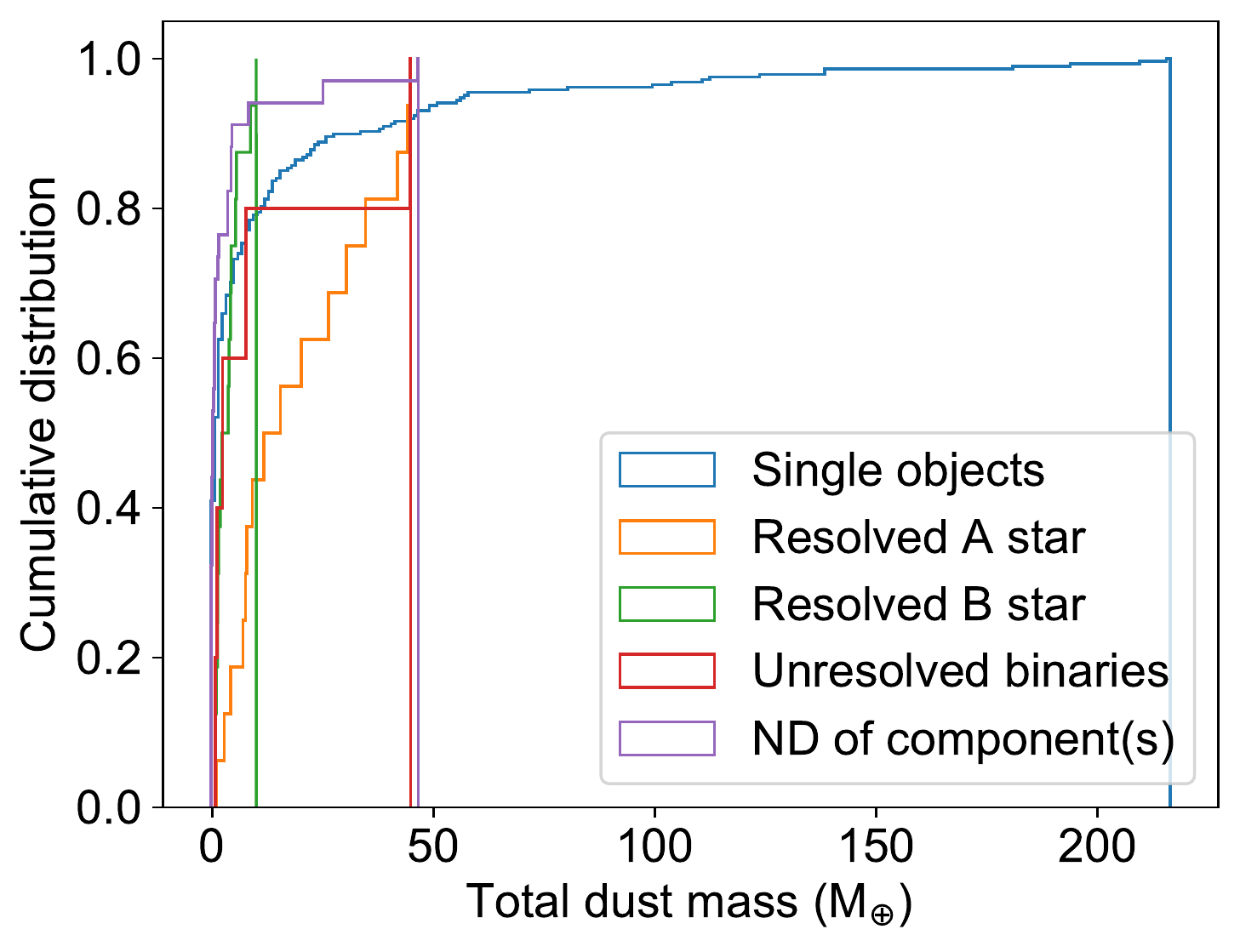}
\caption{{\it Left}: Cumulative distribution of the average mass of the dust around each star in the systems for the two combined samples of Lupus and Ophiuchus, single vs multiple systems are shown. The distributions are very similar up to $\sim$50 M$_{\oplus}$, but discs around single stars can reach dust masses as large as 200 M$_\oplus$. {\it Right}: Cumulative distribution of the total mass of the dust for the different categories of systems}
\label{f:cum_single_com}
\end{center}
\end{figure*}

The distributions of dust disc masses for single and multiple systems are very similar up to M$_{\rm dust}$ $\simeq$ 50 M$_{\oplus}$ (Figure~\ref{f:cum_single}). This result is the same if we include the Ophiuchus discs, as shown in Figure~\ref{f:cum_single_com} (left panel). { A Kolmogorov-Smirnov test comparing the distributions of singles versus multiples gives a p-value of 40\% for Lupus and 51\% for Lupus and Ophiuchus combined. These high p-values suggest that the distributions are nearly identical.} The corresponding cumulative probability distribution for each category is displayed in Figure~\ref{f:cum_single_com} (right panel). The larger sample size strengthens the statistical significance. Discs around single stars may have massive discs up to M$_{\rm dust}$ $\simeq$ 200 M$_{\oplus}$, as derived from the same cumulative distribution. Note that the most massive discs of both Lupus and Ophiuchus have M$_{\rm dust}$ $\simeq$ 200 M$_{\oplus}$.

\begin{figure}
\begin{center}
\includegraphics[width=\columnwidth]{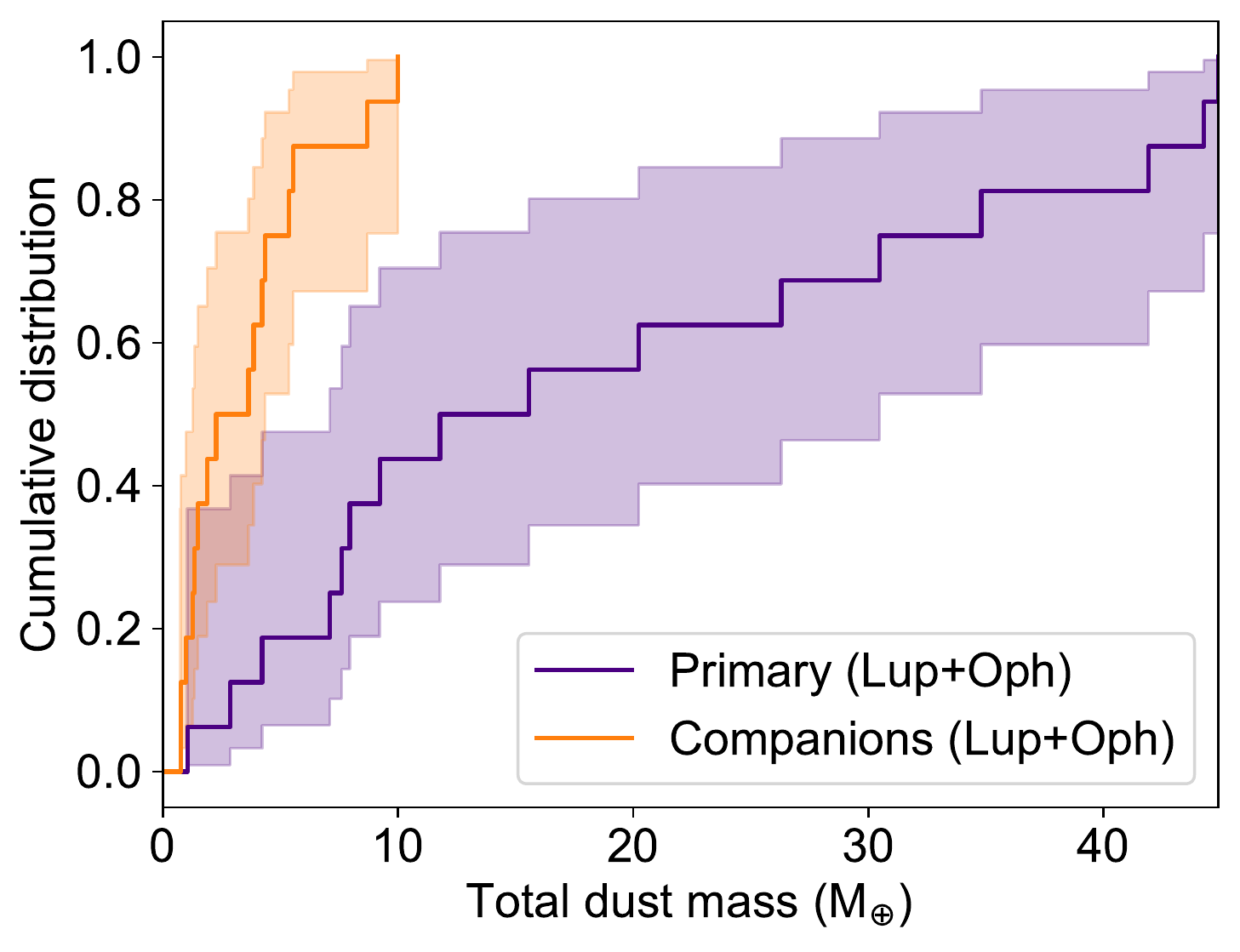}
\caption{Cumulative distribution of the total mass of the dust in the primaries vs the companions as measured for both Lupus and Ophiuchus samples.  Primary stars have more massive discs than their secondaries due to the strong dependence of disc mass on stellar mass and the fact that their discs have larger truncation radii.}
\label{f:cum_a_com}
\end{center}
\end{figure}

Only 4 multiple systems in Lupus have all the components detected also by ALMA. For these systems, the discs around the primaries are more massive than the discs around the secondaries (Figure~\ref{f:cum_a}). This result is confirmed in the larger combined sample of Lupus and Ophiuchus: in Figure~\ref{f:cum_a_com} we show the cumulative distributions of the disc masses for primaries and companions. Discs around the primary stars are indeed systematically more massive than the ones around the secondary stars. This result is likely to be due to the combination of two factors: the strong dependence of disc mass on stellar mass \citep[e.g.,][]{Andrews2013} and the fact that secondaries are expected to have smaller truncation radii. This result is also in agreement with the predictions of \citet{2018MNRAS.473.5630R} that discs around secondary stars have shorter lifetimes than discs around primaries.

\begin{figure}
\begin{center}
\includegraphics[width=\columnwidth]{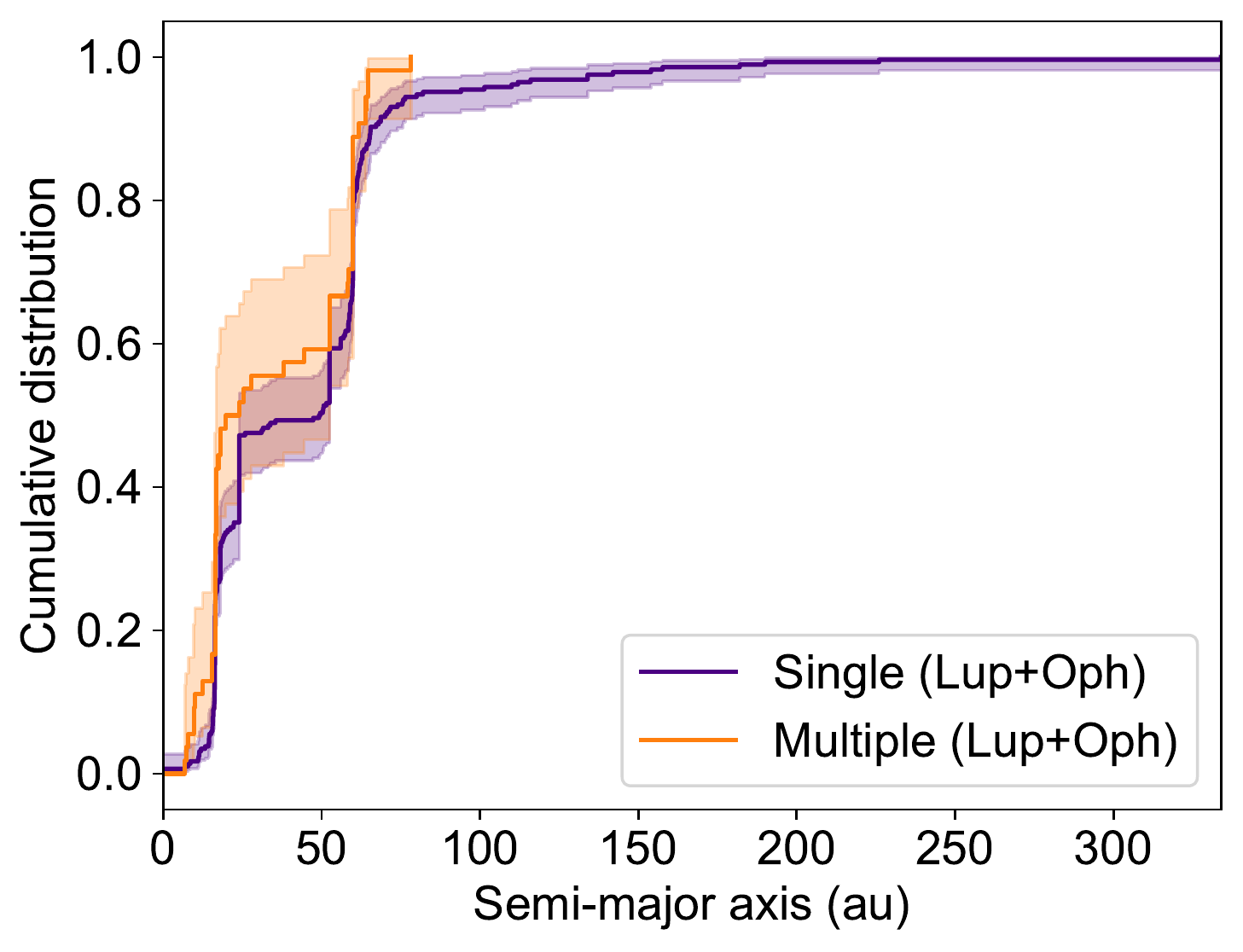}
\caption{Cumulative distribution of the projected semi-major axis of the discs in Lupus and Ophiuchus. The distributions are very similar up to $\sim$80 au, but discs around single stars can be significantly larger.}
\label{f:cum_semi_com}
\end{center}
\end{figure}

As for the mass, the disc size is also influenced by the effects of multiplicity: the semi-major axis is smaller around discs in multiple systems, as seen in Taurus \citep{2019A&A...628A..95M}. From the complete samples of all the discs of Lupus and Ophiuchus we can constrain the maximum size for multiple-system discs to be $\sim$80 au, while discs around single stars can be more than 3 times larger. The semi-major axis cumulative distribution is presented in Figure~\ref{f:cum_semi_com}.

\begin{figure*}
\centering
  \includegraphics[height=0.35\textwidth]{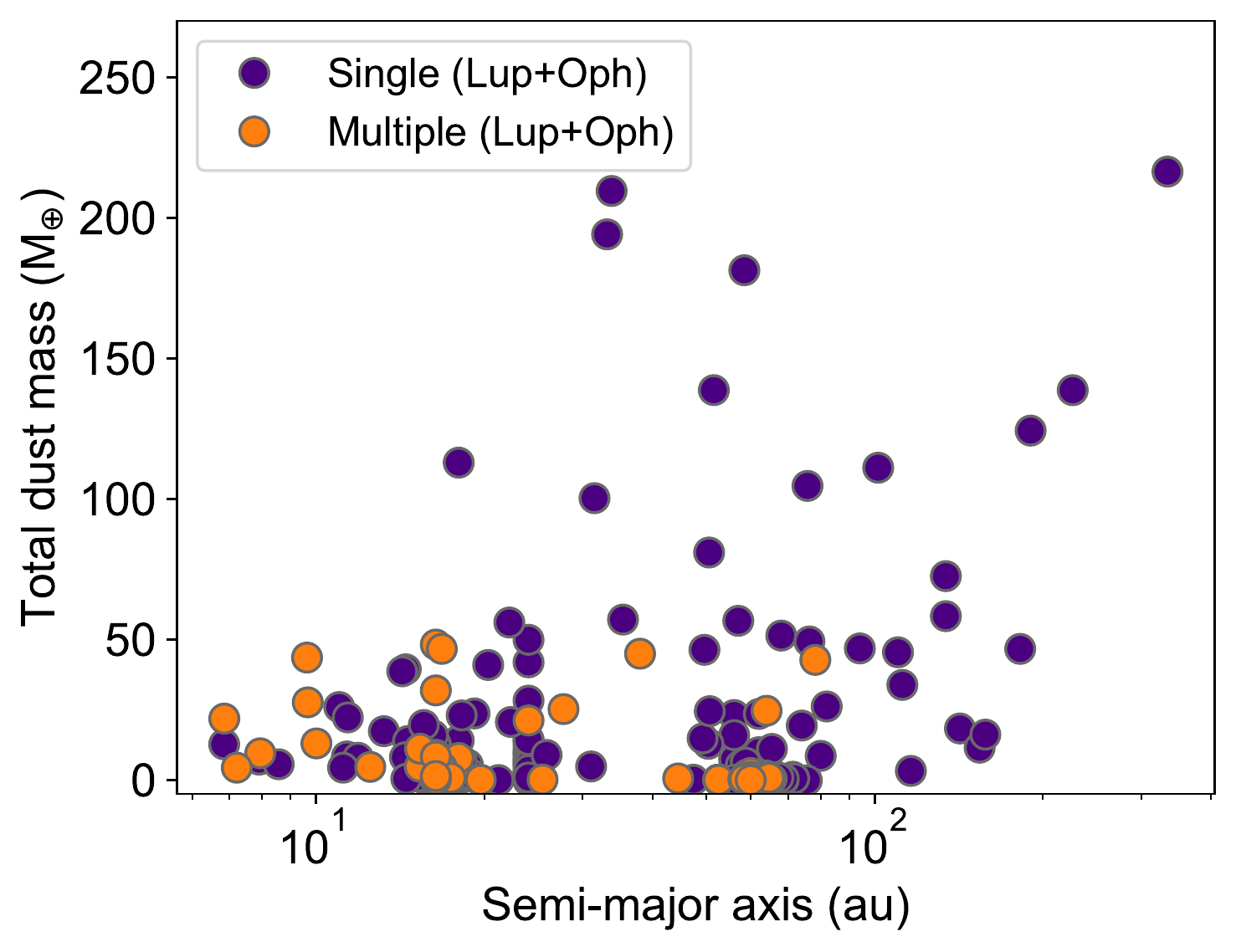} \hfill
  \includegraphics[height=0.35\textwidth]{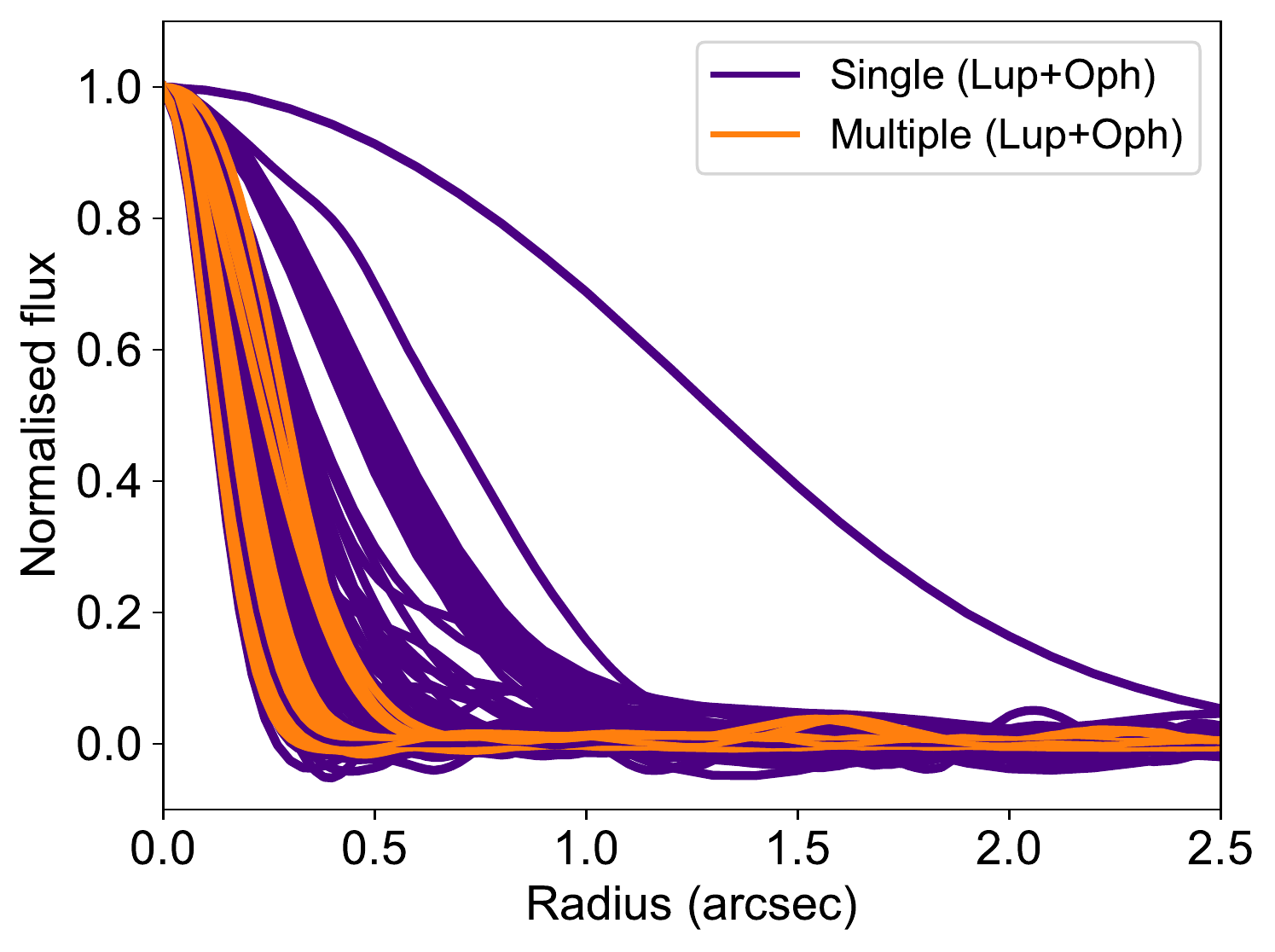}
  \caption{{\it Left}: Dust mass vs semi-major axis of the discs of Lupus and Ophiuchus. Discs around single stars span a larger range in disc sizes and masses.
{\it Right}: Deprojected radial profiles of normalized brightness for discs with peak $SNR > 10$ in Ophiuchus and Lupus. Discs in multiple systems tend to be much more compact than those around single stars.}
\label{f:rad}
\end{figure*}

In Figure~\ref{f:rad} (left panel), we show total disc dust mass as a function of semi-major axis (defined as the radius containing 90$\%$ of the flux) for single and multiple systems. The figure shows, not only that single systems span a larger range in disc sizes and masses, but also that singles have a wider range of masses for a given disc size. This finding suggests that disc around binary systems might have different surface density profiles.  Deriving accurate surface density profiles requires radiative transfer modeling, which is outside the scope of this paper. However, at 1.3 mm the brightness profile is well correlated with the distribution of mass as a function of radius.

In Figure~\ref{f:rad} (right panel), we show the normalized brightness profiles for discs in single and multiple systems, deprojected using the inclination and position angle from {\sc imfit} \citep[from CASA,][]{2007ASPC..376..127M}, for individual sources in Ophiuchus and Lupus. Using the data from both star-forming regions, we obtain the radial profiles for sources where the peak SNR is $>$ 10$\times$rms and the {\sc imfit} gaussian coincides with the signal peak, thereby excluding transition discs as well as poorly resolved sources. The figure includes 147 radial profiles between single and binary stars for both samples. Discs in multiple systems are clearly more compact than single star discs. The same finding is presented in \cite{2019A&A...628A..95M} for the Taurus region.

%
%

Even if the influence of multiplicity is evident on the mass and size of the protoplanetary discs, \cite{2007A&A...462..345D} found that the properties of exoplanets around single and wide binaries are similar, with the exception of a possible effect on the eccentricity of the planet orbits. For tight binary systems ($<$ 40 au) the disc life time is expected to be so short that planets would have to form within 1 Myr \citep{2012ApJ...745...19K,2015ApJ...813...83C}. The fraction of binary systems in Lupus and Ophiuchus is only 4\% from the NIR-Lupus and NIR-ODISEA surveys, even if they are not sensitive to very short-period binaries due to the NACO spatial resolution (0\farcs1), corresponding to { $\sim$20 au} projected separation { at 150 pc}.

\section{Summary and conclusions}
\label{s:con}
We have conducted a NIR AO imaging survey of all the Class II and III objects of the Lupus star-forming region. We complemented our VLT/NACO L$^{\prime}$ band observations of 47 stars with archival data to cover almost all the 125 circumstellar discs of Lupus. Only 20 faint objects were not included in this analysis. { Our study is sensitive to visual binaries with projected separations between 20--4800 au ---spectroscopic, tight, and large separation binaries are not included in our statistics. } In total we report the detection of { 11 potentially bound} binary systems and { 2} triple systems, where the primaries have different spectral type spanning from A to L. The frequency of visual companion multiplicity in Lupus { in the range 20-4800 au is then 12\% (13/105)}. Among the companions, { one object is} in the planetary regime (12 \MJup), but we cannot exclude { it as a} background contaminant with the available data.

To strengthen the statistical analysis, we combined this Lupus survey with its twin in Ophiuchus \citep{Zurlo2020}. The results that were obtained with the combined 341 (105+236) PMS stars are listed below. { The following statistics are applicable to visual companions in the range of projected separations $\sim$20--4000 au and flux ratios $>$ 0.01:}

\begin{enumerate}
\item Considering the Lupus and Ophiuchus regions, the fraction of stellar multiplicity for circumstellar discs with { visual companions (separations $\sim$20--4000 au, flux ratios $>$ 0.01) is 16\%}. 
  
\item The distribution of the dust mass of discs around single stars and multiple systems are similar up to M$_{\rm dust}$ = 50 M$_{\oplus}$. Higher mass discs (up to $\sim$200 M$_{\oplus}$) are only found around single stars. 

\item Discs around primary stars are systematically more massive than the ones around their companions. 

\item Discs around single stars can have semi-major axes up to 300 au, while discs around multiple systems are 3 times smaller. 

\item Discs around stars in multiple systems have flux distributions 
that tend to be much more compact than those seen for single stars. This suggests significant differences in their surface density profiles. 

\item The visual binaries identified in our AO IR survey have separations $>${ 20} au and mostly affect discs properties in the 10$\%$ upper end of the mass and size distributions. The incidence of tight visual binaries (with separations of 10-40 au) that are likely to prevent the formation of planets is rather low (4 \%), possibly due to the limited angular resolution of our survey.

\end{enumerate}

\section*{Acknowledgements}
{ The authors are grateful to the anonymous referee for their comments that improved the quality of the manuscript.} A.Z. acknowledges support from the FONDECYT {\it Iniciaci\'on en investigaci\'on} project number 11190837. L.C. acknowledges support from FONDECYT Regular number 1171246. V.C. is grateful for Australian Research Council funding  via  DP180104235. S.P. acknowledges support from FONDECYT grant number 1191934 and the Joint Committee of ESO and the Government of Chile. J.P.W. acknowledges support from the National Science Foundation under grant AST-1907486. R.C. acknowledges financial support from CONICYT Doctorado Nacional N$^\circ$\,21161487 and CONICYT PIA ACT172033. This paper makes use of the following ALMA data: ADS/JAO.ALMA 2015.1.00222.S. and 2018.1.00437.S. ALMA is a partnership of ESO (representing its member states), NSF (USA) and NINS (Japan), together with NRC (Canada), MOST and ASIAA (Taiwan), and KASI (Republic of Korea), in cooperation with the Republic of Chile. The Joint ALMA Observatory is operated by ESO, AUI/NRAO and NAOJ. Based on ESO observations programs number 073.C-0379, 085.C-0012, 095.C-0610, 097.C-0572, 0103.C-0466. \\

\section*{Data availability}
The data underlying this article are available in the ESO archive at https://archive.eso.org/eso/eso\_archive\_main.html, and can be accessed with program numbers 073.C-0379, 085.C-0012, 095.C-0610, 097.C-0572, 0103.C-0466.




\bibliographystyle{mnras}
\bibliography{bin_lupus} 


\appendix

\section{Not observable Lupus objects}

\begin{table}
\caption{{List of the not observable Lupus objects in the NIR with NACO, then excluded in the statistical analysis.}} 
\label{t:non_o}
\centering
\begin{tabular}{lcc}
\hline
\hline
Name & RA & DEC  \\
\hline
SSTc2dJ154301.3-340915 & 	15:43:01.29 & 	-34:09:15.40 \\
SSTc2dJ154302.3-344406 & 	15:43:02.29 & 	-34:44:06.20 \\
SSTc2dJ160115.6-415235 & 	16:01:15.55 & 	-41:52:35.30 \\
SSTc2dJ160703.9-391112 & 	16:07:03.85 & 	-39:11:11.60 \\
SSTc2dJ160714.0-385238 & 	16:07:14.00 & 	-38:52:37.90 \\
SSTc2dJ160754.7-391545 & 	16:07:54.75 & 	-39:15:44.60 \\
SSTc2dJ160801.7-391231 & 	16:08:01.75 & 	-39:12:31.60 \\
SSTc2dJ160815.0-385715 & 	16:08:14.96 & 	-38:57:14.50 \\
SSTc2dJ160828.1-391310 & 	16:08:28.10 & 	-39:13:10.00 \\
SSTc2dJ160831.1-385600 & 	16:08:31.10 & 	-38:56:00.00 \\
SSTc2dJ160851.4-390530 & 	16:08:51.43 & 	-39:05:30.40 \\
SSTc2dJ160858.3-390749 & 	16:08:58.30 & 	-39:07:49.40 \\
SSTc2dJ160916.4-390444 & 	16:09:16.43 & 	-39:04:43.70 \\
SSTc2dJ160920.3-390402 & 	16:09:20.30 & 	-39:04:01.60 \\
SSTc2dJ160923.2-390407 & 	16:09:23.15 & 	-39:04:07.40 \\
SSTc2dJ160934.2-391513 & 	16:09:34.18 & 	-39:15:12.70 \\
SSTc2dJ160939.3-390432 & 	16:09:39.29 & 	-39:04:31.80 \\
SSTc2dJ161013.1-384617 & 	16:10:13.06 & 	-38:46:16.80 \\
SSTc2dJ161027.4-390230 & 	16:10:27.43 & 	-39:02:30.20 \\
SSTc2dJ161204.5-380959 & 	16:12:04.48 & 	-38:09:59.00 \\
\hline
\end{tabular}
\end{table}

\bsp	
\label{lastpage}
\end{document}